\documentclass[a4paper,11pt]{article}
\usepackage{jinstpub} % for details on the use of the package, please see the JINST-author-manual
\usepackage{lineno}
\usepackage{epsfig}
\usepackage{array,tabularx,epsfig,mathrsfs,graphicx,rotating}
\usepackage{ifthen}
\usepackage{comment}
\usepackage{ragged2e}
\PassOptionsToPackage{hyphens}{url}
\usepackage[hyphens]{url}
\usepackage{hyperref}
\usepackage{listings}
\usepackage{epstopdf}
\usepackage{color}
\usepackage{float}
\usepackage{subfig}

\hypersetup{
  colorlinks=true,
  linkcolor=blue,
  citecolor=blue,
  urlcolor=blue
}

\graphicspath{{figs/}}

\newcommand{\beq}{\begin{equation}}
\newcommand{\eeq}{\end{equation}}

\newcommand{\geant}{GEANT4}
\newcommand{\cere}{ Cherenkov}

\chardef\til=126

\definecolor{mygreen}{rgb}{0,0.6,0} \definecolor{mygray}{rgb}{0.5,0.5,0.5} \definecolor{mymauve}{rgb}{0.58,0,0.82}

\title{\boldmath
Geant4 simulations of sampling and homogeneous 
hadronic calorimeters with dual readout for future colliders}

\author[a]{S.V.~Chekanov}

\author[b]{S.~Eno}

\author[a]{S.~Magill}
\affiliation[a]{HEP Division, Argonne National Laboratory,
9700 S.~Cass Avenue,
Lemont, IL 60439, USA.}

\author[b]{C.~Palmer}

\author[b]{L.~Wu}

\affiliation[b]{Dept. Physics, U. Maryland, College Park, MD, 20742, USA}

\note{Preprint ANL-HEP-186226}

\abstract{
Hadronic calorimeters with dual readout measure both scintillation and Cherenkov lights produced in their  active media. They offer improvements in energy resolution and, therefore,
have become increasingly interesting due to the need for precision jet measurements at Higgs factories. 
This paper presents
\geant\ simulations of single-particle responses in sampling and homogeneous  calorimeters, and demonstrates the effect of
inclusion of Cherenkov light in the reconstruction of energies. The simulations 
are performed with a single-photon precision. 
}


\begin{document}
\maketitle
\flushbottom
%%%%%%%%%%%%%%%%%%%%%%%%%%%%%%%%%%%%%%%%%%%%%%%%%%%%%%%%%%%%%%%%%%

Dual readout hadronic calorimeters, which simultaneously detect scintillation and Cherenkov light, 
provide an opportunity to improve the resolution of  hadrons and jets.
A typical hadronic calorimeter (HCAL) has a sampling term of about $50\%/\sqrt{E}$ for jets, where $E$ is the energy. Algorithms
based on particle flow can reduce the sampling term to about $30\%/\sqrt{E}$~\cite{Sefkow:2015hna}. 
Dual readout~\cite{Magill_2012,AKCHURIN2005537,Lee:2017xss, Pezzotti:2022ndj,Lucchini:2020bac} can reduce the sampling term  to  10-20\%. 
The dual readout technique can be applied to homogeneous crystal calorimeters~\cite{Magill_2012,Lucchini:2020bac,Lee:2017xss},
fiber-based sampling calorimeters~\cite{Lee:2017xss} and a sandwich-style calorimeter \cite{Takeshita:2023fap}.
Experimentally, the  reduction of the sampling term has been  demonstrated by the DREAM collaboration for  fiber-based sampling calorimeters~\cite{Antonello:2018sna}.

Calorimeters with a sandwich-like structure, where passive absorber layers  interleave with active layers, is a popular technology used in many past -- and also in the  conceptional designs of future  hadronic calorimeters. 
A popular modern version is the so-called ``high granularity'' calorimeter.
Numerous simulations of such detectors   for lepton colliders exist, and 
both simulations and prototypes have been explored by the CALICE collaboration~\cite{CALICE:2012ami}. Historically, the detector collaborations for the TESLA linear $e^+e^-$ collider  envisioned a lead-scintillator sandwich
calorimeter with a shashlik readout~\cite{articleTESLA}.
Later, the ILC project investigated several options for finely-segmented sampling calorimeters based on a sandwich-style design~\cite{SCHUWALOW2009258}, although   scintillating-fiber calorimeters with dual readout were also considered by the IDEA collaboration~\cite{Bedeschi:2021nln} as an option.
The CLIC and FCC-ee projects also consider finely-segmented sampling calorimeters~\cite{Viazlo_2019}  in which  the tungsten or steel absorbers are  sandwiched between the scintillation material. 
More recently, this calorimeter design was adopted by the CLD detector \cite{Bacchetta:2019fmz} 
for the FCC-ee project.
No dual readout was assumed for the baseline designs.

This paper explores dual readout for the traditional HCAL design with a sandwich-like structure, as considered for the TESLA, ILC, CLIC, CEPC, and FCC-ee  $e^+e^-$ projects, using Monte Carlo simulations.
Such designs are compared with a traditional homogeneous calorimeter
made of PbWO4 crystals and some variations of the sampling calorimeters. 

\section{\geant\ simulations}

The effect of Cherenkov light on measured energy resolutions
in dual readout systems significantly depends on the photocollecting technology and other instrumental effects.
In this paper we will be interested in the {\em physics principles} of such systems as predicted by \geant\ simulations assuming the photocollection
efficiency is 100\%.

The \geant\ (version 10-07) \cite{ALLISON2016186} simulations of optical photons
are very CPU consuming. Therefore, the simulations were performed for
a single calorimeter ``tower'' for each of the design choices described later.
Each  tower has a size of about 5.6 nuclear interaction
lengths ($\lambda_I$) in the $Z$ (longitudinal) direction 
in order to guarantee $>95\%$ shower containment in the longitudinal 
direction for single-particle energies
between 1 and 40\,GeV.
The transverse ($X-Y$) size of each towers is $1\lambda_I$, which provides
more than $70\%$ lateral containment for this energy range. 
This transverse size is close to the sizes of typical 
towers used in the contemporary calorimeters.  Our decision to simulate  
calorimeter towers, instead of groups of towers which can fully contain
the hadronic shower, was also motivated by the desire to reduce the CPU usage of the simulations.
All calorimeter towers used in the simulation are not compensating,
i.e. the responses to the electromagnetic ($e$) and  hadronic ($h$)
shower components are not the same, i.e. $h/e<1$.

As a test that is discussed later,  towers were also simulated with a
$4\lambda_I$ width in $X-Y$, providing 
a $2\lambda_I$ radius for the transverse shower, leading to about $95\%$ lateral shower containment. Such checks were performed with a reduced precision due to  CPU limitations.

The  simulation of the calorimeter response
for inelastic processes is based on the  FTFP\_BERT physics list. 
Both the scintillation and Cherenkov physics processes were enabled in the simulations.
The geometry description of the towers were  implemented in the {\sc DD4HEP} package \cite{dd4hep} which provides  a flexible description of simulation geometries.

The \geant\ simulations were used to create 3000  
events of each particle type, i.e. electrons ($e^-$),
pions ($\pi^-$), protons ($p$) and neutrons ($n$).
The simulations of scintillation ($S$) and Cherenkov ($C$) photons are performed without
approximations (such as an artificial decrease of the scintillation yield, or artificial removal of photons in the stepping action) in order to reduce the rate of simulated optical photons and to 
reduce the computer CPU. In this sense, the simulations are performed
with the ``single-photon'' precision since the ($x,y,z$) position and wavelength of each photon were recorded and stored. This is important for the final analysis of optical photons in various active layers of the sampling calorimeter.
The simulations also produce positions and particle types for the energy hits, which are
traditionally  used in energy reconstruction. This information, however, was not used in this analysis since
the primary focus of this paper is the distributions of the optical photons
from the simulations.

In the following, the 
$S$  and $C$ symbols represent the light yields (i.e. the number of optical photons) calibrated using the single electron response. In this calibration procedure the total numbers of scintillation or Cherenkov photons for a given incoming particle
were divided by the  average number of photons obtained for single electrons with the same energy. Thus for this definition, $<S>=<C>=1$ for electron collisions. 

%%%%%%%%%%%%%%%%%%%%%%%%%%%%%%%
\section{Designs of the calorimeter towers}
%%%%%%%%%%%%%%%%%%%%%%%%%%%%%%%
\label{sampling}

\subsection{40 layer sampling calorimeter (40L-PFQ)}
\label{sampling}
We started with a geometry for a single tower 
based on the existing designs
of the sampling calorimeters of the ILC/CLIC detectors,
with alternating active and passive
absorber layers. 
The tower consists of 40 layers or plates. Each layer has an absorber (Fe) with the width of 
1.8\,cm. The two active media, positioned after the steel plate, 
are  polystyrene (for $S$ light) and quartz (for $C$ light). These two layers  are separated by  0.1\,cm of steel. 
Each active media has a width of 0.5\,cm. The final part of this design is the steel plate of 0.1\,cm width that protects the  quartz layer and forms the wall of the single layer. 
This results in  2\,cm of absorber and 1\,cm  of active media for each of the 40 layers.
This is similar to the  CLIC and ILC-style HCAL designs, where the active media 
is typically polystyrene (for scintillation sampling calorimeter) or resistive plate chambers.
The interaction length of the $40\times (1.8+0.1+0.1$\,cm)  steel absorbers  is 
4.77~$\lambda_I$. When adding the active media, the total  interaction length of the tower is 
5.48~$\lambda_I$.
The transverse size of the tower was $20\times 20$\,cm.
The simulation uses the scintillation efficiency of 10 photons/keV, typical for plastic scintillators manufactured by Eljen Techology~\cite{EJefficiency}. Scintillation and Cherenkov signals were counted in
polystyrene and quartz layers, respectively. 
The $C$ light created by polystyrene was disregarded for the final calculations of the light yields. Quartz does not produce any observable  $S$ signal.

The above design, which will be denoted 40L-PFQ (40 layers made of polystyrene, steal and quartz),  provides a good containment of  hadronic showers for single particles up to 20\,GeV in  the longitudinal direction. The lateral shower containment, however, is about 70-80\%
depending on the incoming energy of particles. This makes the design of this HCAL 
tower  non-compensating, even if  it was originally designed to be compensating assuming minimal leakage. 
Though this design does not assume high granularity in the transverse direction, which is often used in conceptual HCAL designs, 
the goal of this simulation is to understand the effect  from dual-readout 
correction when $C$ light is added to the $S$ light with the goal
to reduce the overall resolution.  

\begin{figure}[htb!]
  \begin{center}
  \includegraphics[width=0.7\textwidth]{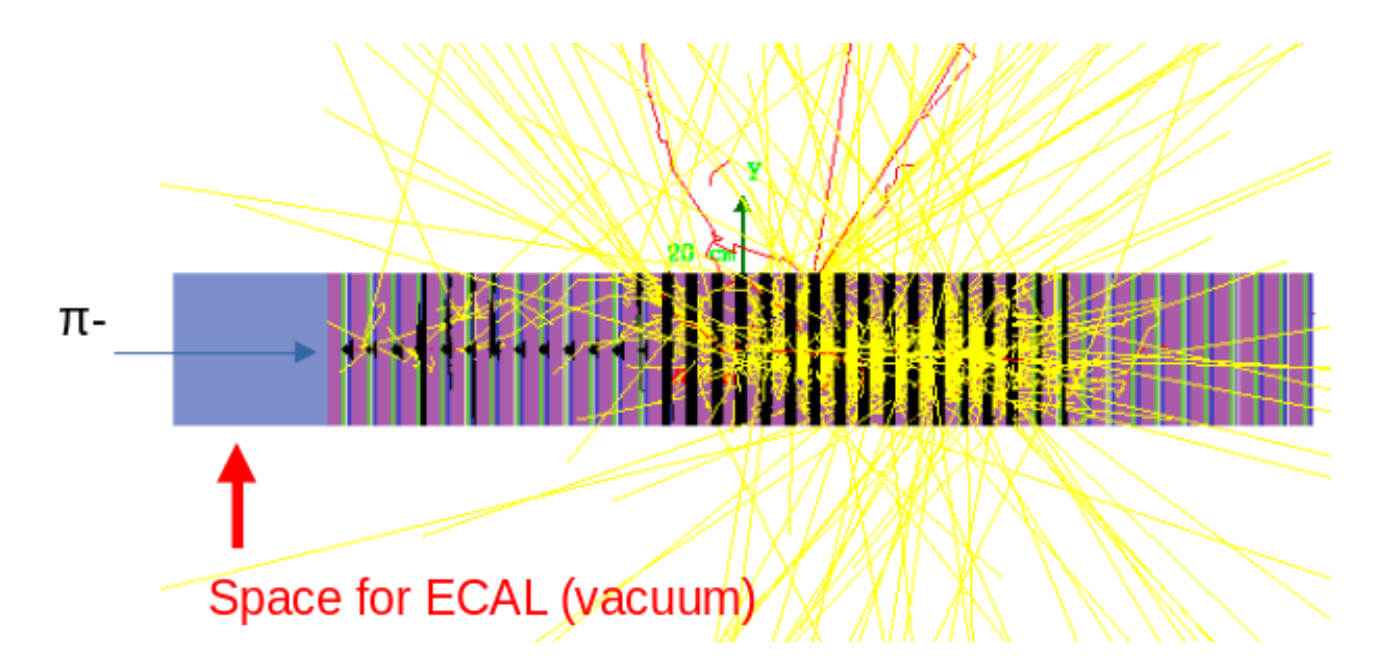}
  \end{center}
  \caption{
  An example of the shower produced by a 10\,GeV pion ($\pi^-$)  interacting in the 40L-PFQ tower. The red colors are the absorber (steel). The green (blue) strips between the absorber are polystyrene (quartz). Black lines are optical photons (which typically fill the space between the absorbers due to their large numbers and many reflections before absorption or detection). The yellow lines are photons, and the red lines are muons.
  }
  \label{fig:hcal_vew}
\end{figure}

\begin{figure}[h]
  \begin{center}
  \includegraphics[width=0.7\textwidth]{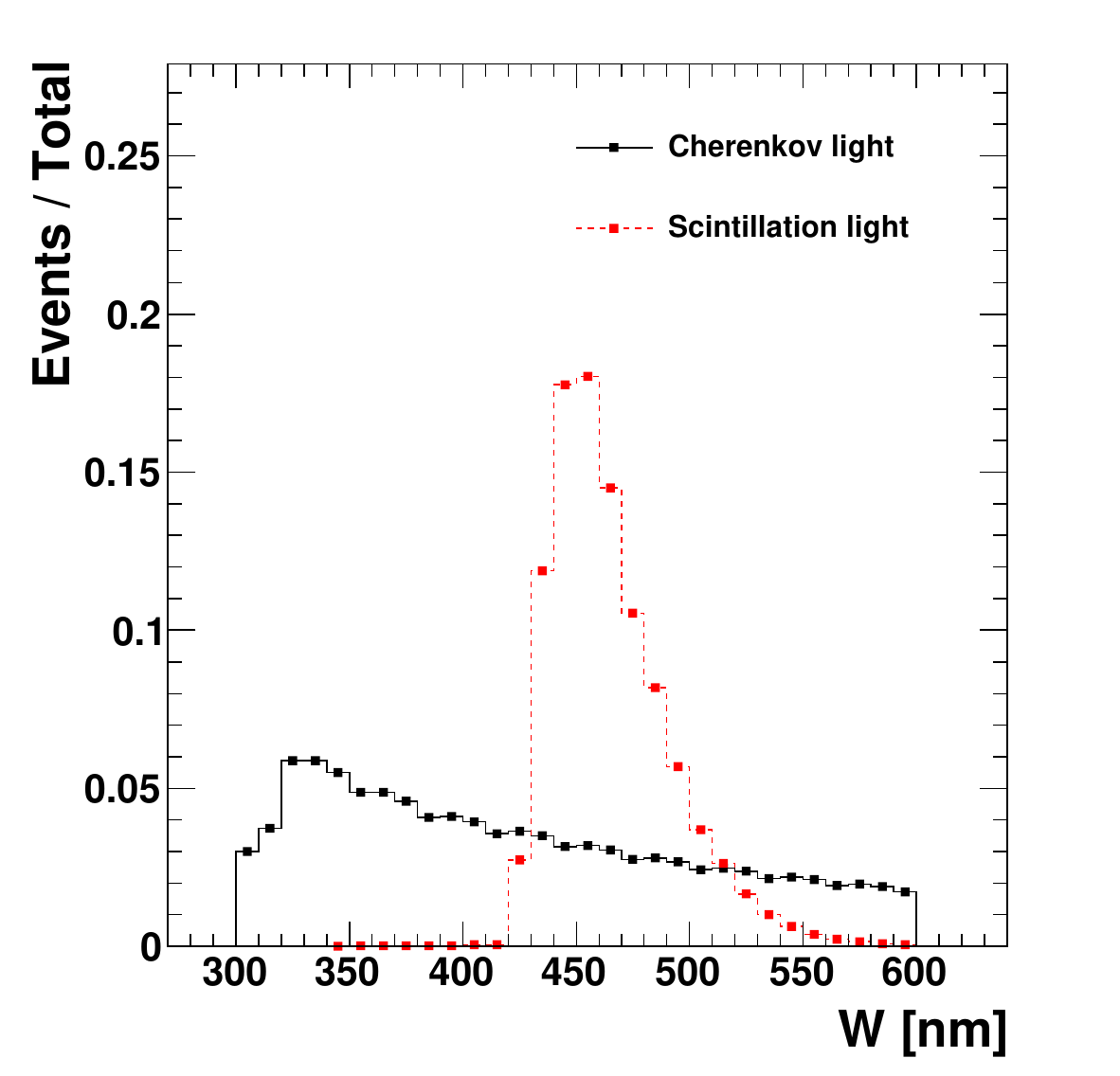}
  \end{center}
  \caption{
 The wavelength distribution obtained from the simulation of the 40L-PFQ sampling hadronic calorimeter with steel absorber and polystyrene and quartz active media. The distributions are normalised to the unit area.
  }
  \label{fig:hcal_view}
\end{figure}

Figure~\ref{fig:hcal_vew} shows a  \geant\ simulation of the response of a single 10\,GeV pion ($\pi^-$) in the tower.
Optical photons are counted in the active media in the wavelength range of 300 -- 600\,nm. No optical filters are used.
The wavelength distributions of scintillation and Cherenkov lights are shown in Fig.~\ref{fig:hcal_view}. The distributions are normalized to unit area.

\begin{figure}[h!]
  \begin{center}
  \includegraphics[width=0.9\textwidth]{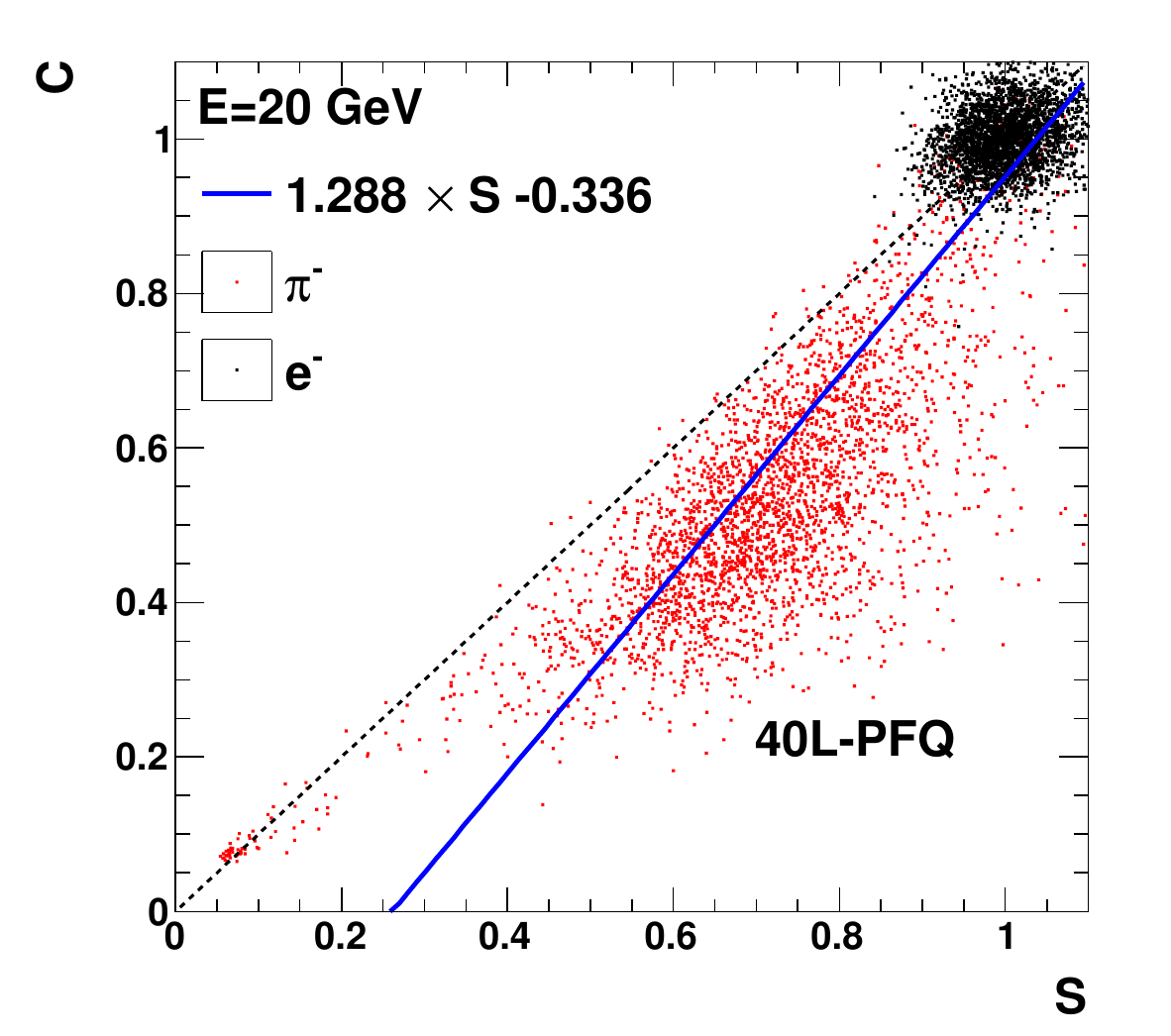}
  \end{center}
  \caption{
 Correlation between calibrated scintillation yield ($S$) and  Cherenkov yield ($C$) for a sandwich-style sampling hadronic calorimeter, 40L-PFQ, with 40 layers of  steel absorber, and with polystyrene and quartz active media. 
 The blue line shows the $\chi^2$ fit of the pion data.
 The signals were calibrated using the 20\,GeV electron beam.
  }
  \label{fig:correlation}
\end{figure}

Figure~\ref{fig:correlation} shows the correlation between the $S$ and $C$ signals
for the 20\,GeV pions. As a reminder, the $S$ and $C$ represent the calibrated yields of optical photons using simulations of electron beams; thus $<S>=<C>=1$ are expected for electron beams. The blue line represents the $\chi^2$ fit to the pion data.
We observe a correlation between $S$ and $C$ photons with a significant
spread of data around the fit line.

The simulations were used to calculate the responses to the
electromagnetic and hadronic shower components, $e$ and $h$.
The simulations were run two times, once for the pion beams, and once for the 
electron beams. Then the ratio of the number of optical photons was taken separately
for scintillation and Cherenkov photons.
The estimated ratio of the hadronic to electromagnetic response are $(h/e)_S = 0.71$ (for the scintillation light) and $(h/e)_C=0.52$ (for Cherenkov light) for a 20\,GeV pion beam, yielding $\kappa \equiv (1- (h/e)_S) /  (1- (h/e)_C)= 0.6$. The energy dependence of the $\kappa$ was
found to be small in the range between 1 and 20\,GeV.
The corrected energies are evaluated \cite{Lee:2017xss} as

\begin{equation}
E=\frac{S- \kappa\cdot C}{1-\kappa},
\label{eqene}
\end{equation}
where $S$ and $C$ are scintillation and Cherenkov light yields after calibration to the signal from electron beams.

\begin{figure}[hbt!]
\begin{center}

   \subfloat[5 GeV] {
   \includegraphics[width=0.48\textwidth]{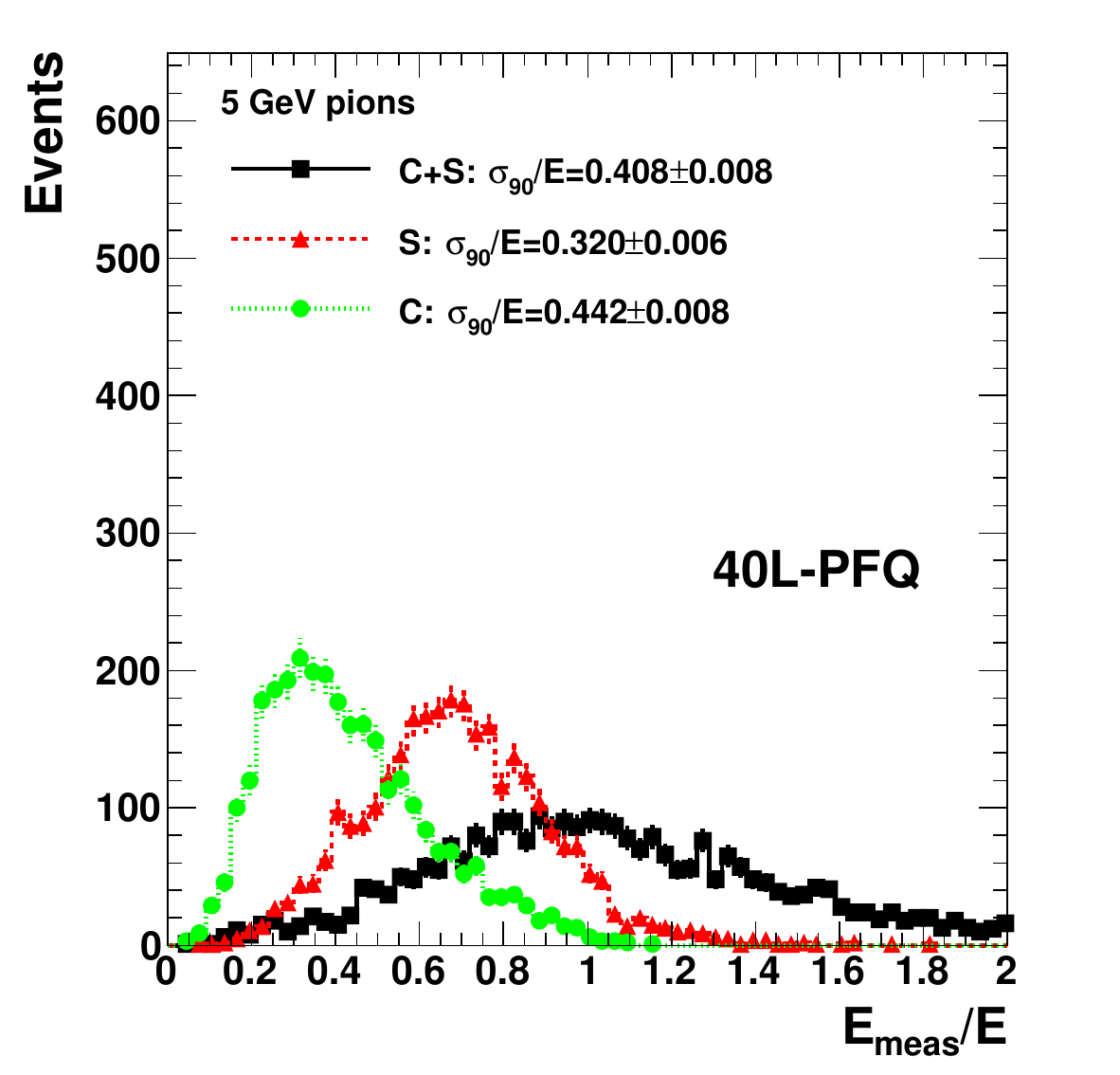}\hfill
   }
   \subfloat[10 GeV] {
   \includegraphics[width=0.48\textwidth]{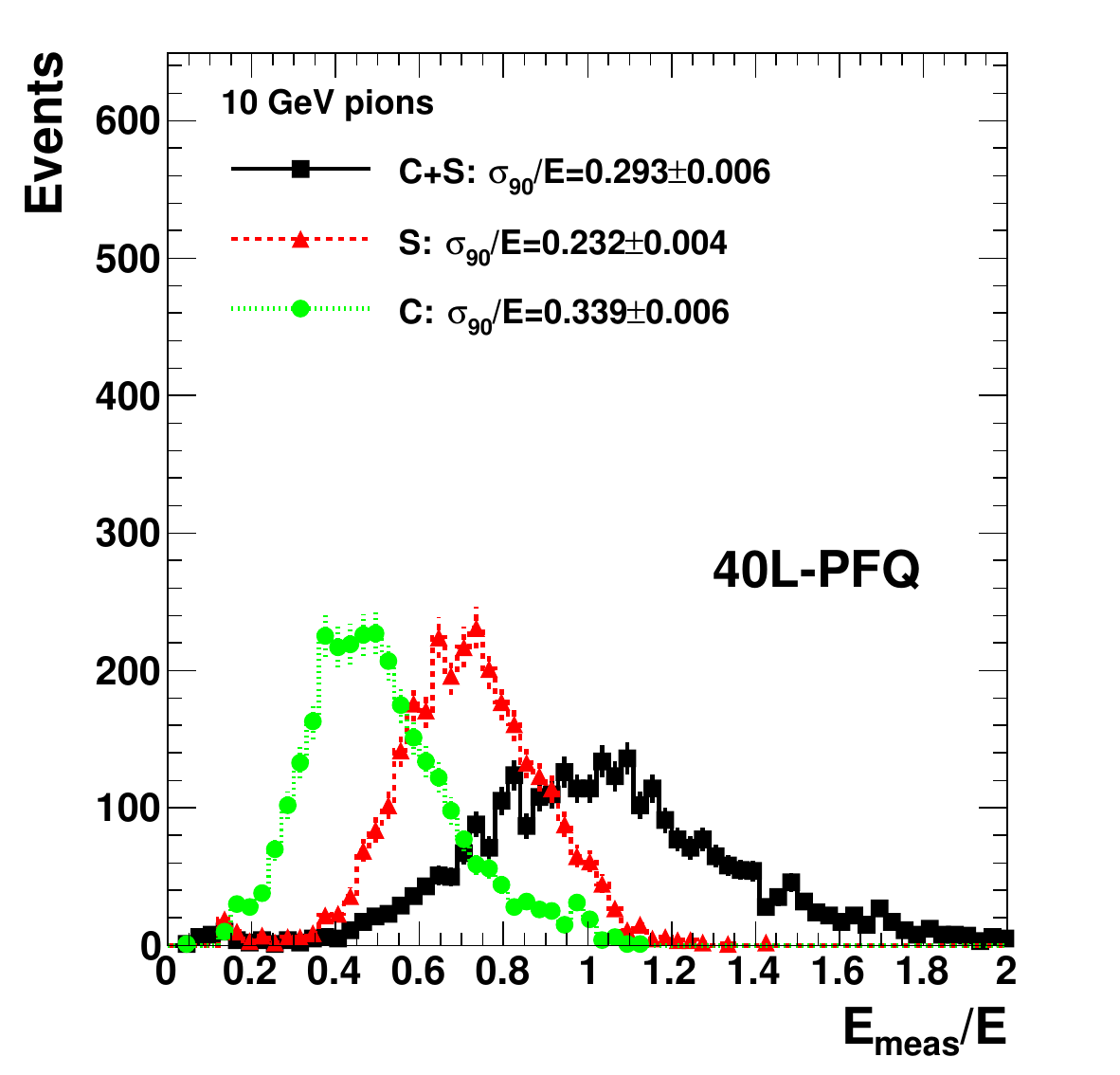}
   }

   \subfloat[20 GeV] {
   \includegraphics[width=0.48\textwidth]{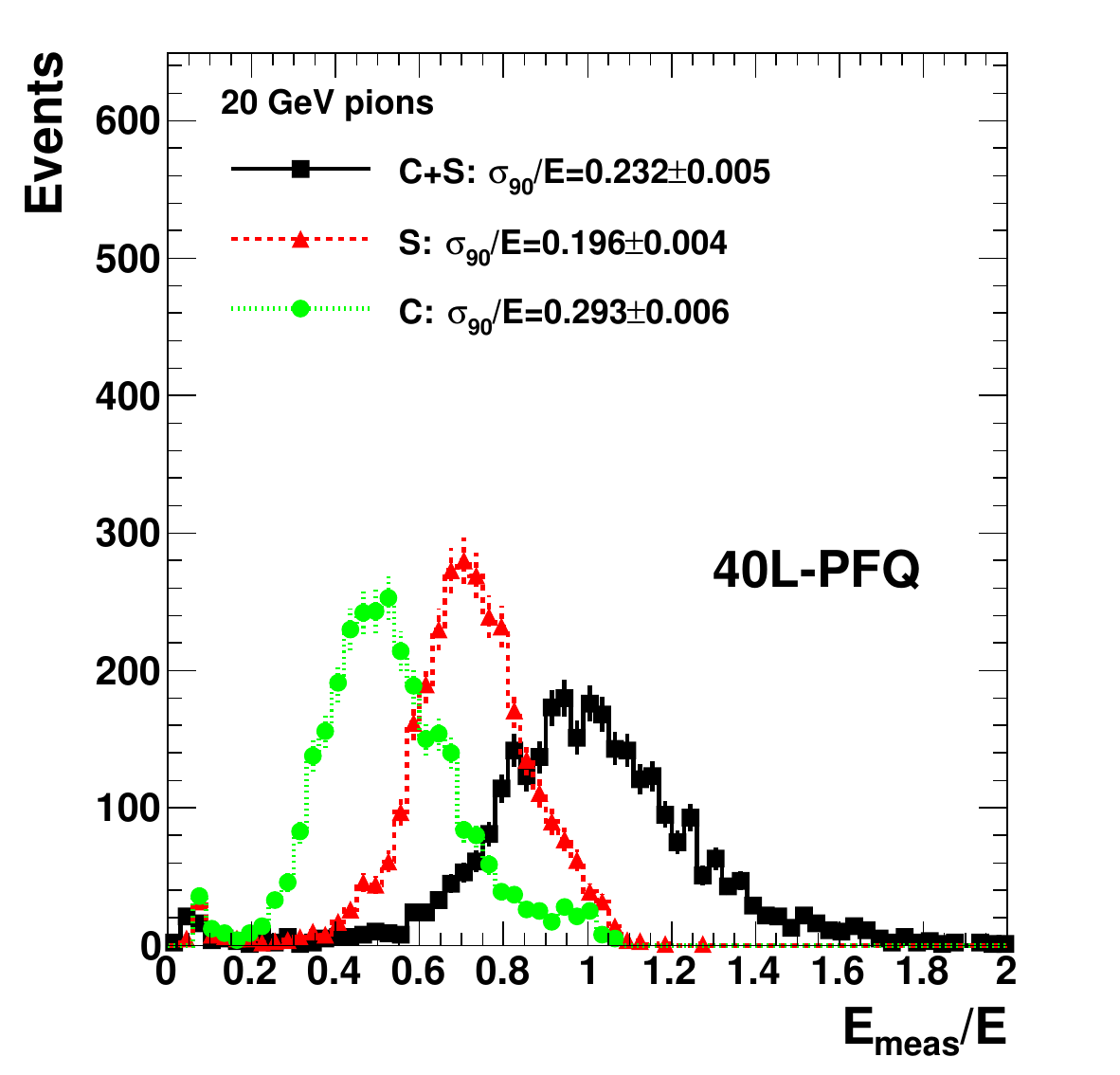}\hfill
   }
   \subfloat[40 GeV] {
   \includegraphics[width=0.48\textwidth]{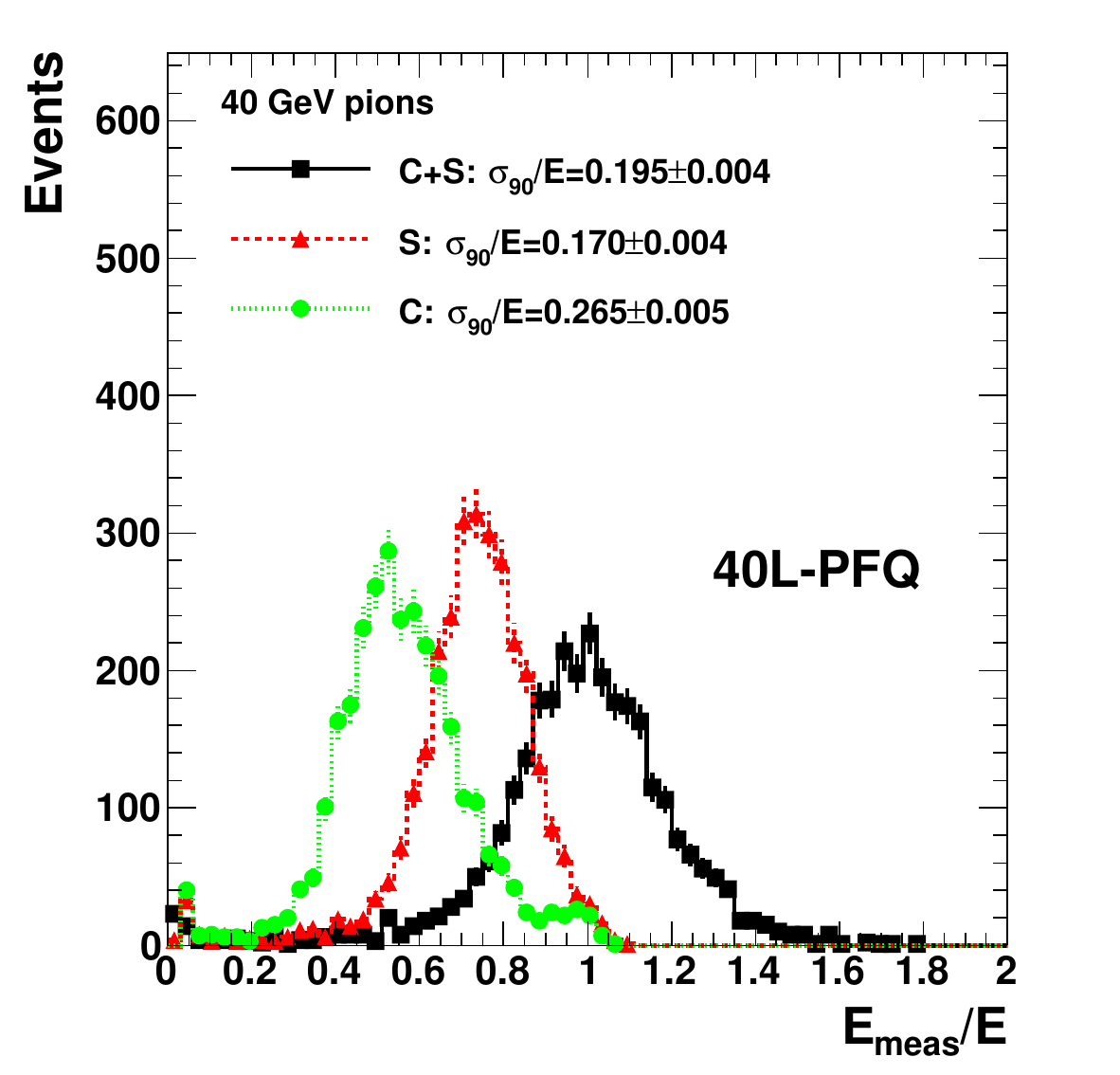}
   }
\end{center}
\caption{
The distributions of the $S$, $C$ and $S+C$ photons for the pion  beams with the
energies 5-40\,GeV for the 40 layers sampling calorimeter. 
The mean and the $\sigma_{90}$ of the signal profiles are also shown.}
\label{fig:corrections1}
\end{figure}

Figure~\ref{fig:corrections1}   shows the corrected
energies of the pions  after applying the dual-readout corrections. The peaks 
were calculated from the mean values. The widths ($\sigma_{90}$) were calculated from the root mean square (RMS) of the central 90\% of the entries in a histogram to
eliminate the effect of the non-Gaussian tails.
As an estimate of the resolution in terms of the RMS, after eliminating the effect
of the long tails, the $\sigma_{90}$ should be multiplied by a factor 1.25 to obtain the RMS \cite{RMS90}.
Compared to the readout based on counting of scintillation photons only,
the improvement in the $\sigma_{90}/E$ values after including $C$ signals is not observed.
To first approximation, the same correction applied to neutrons also shows no improvements compared to $S$ photons. Similar results have been obtained for  kaons and protons (not shown).

\subsection{Homogeneous crystal calorimeter (5L-PBWO4)}
\label{crystal}
\begin{figure}[hbt!]
  \begin{center}
  \includegraphics[width=0.9\textwidth]{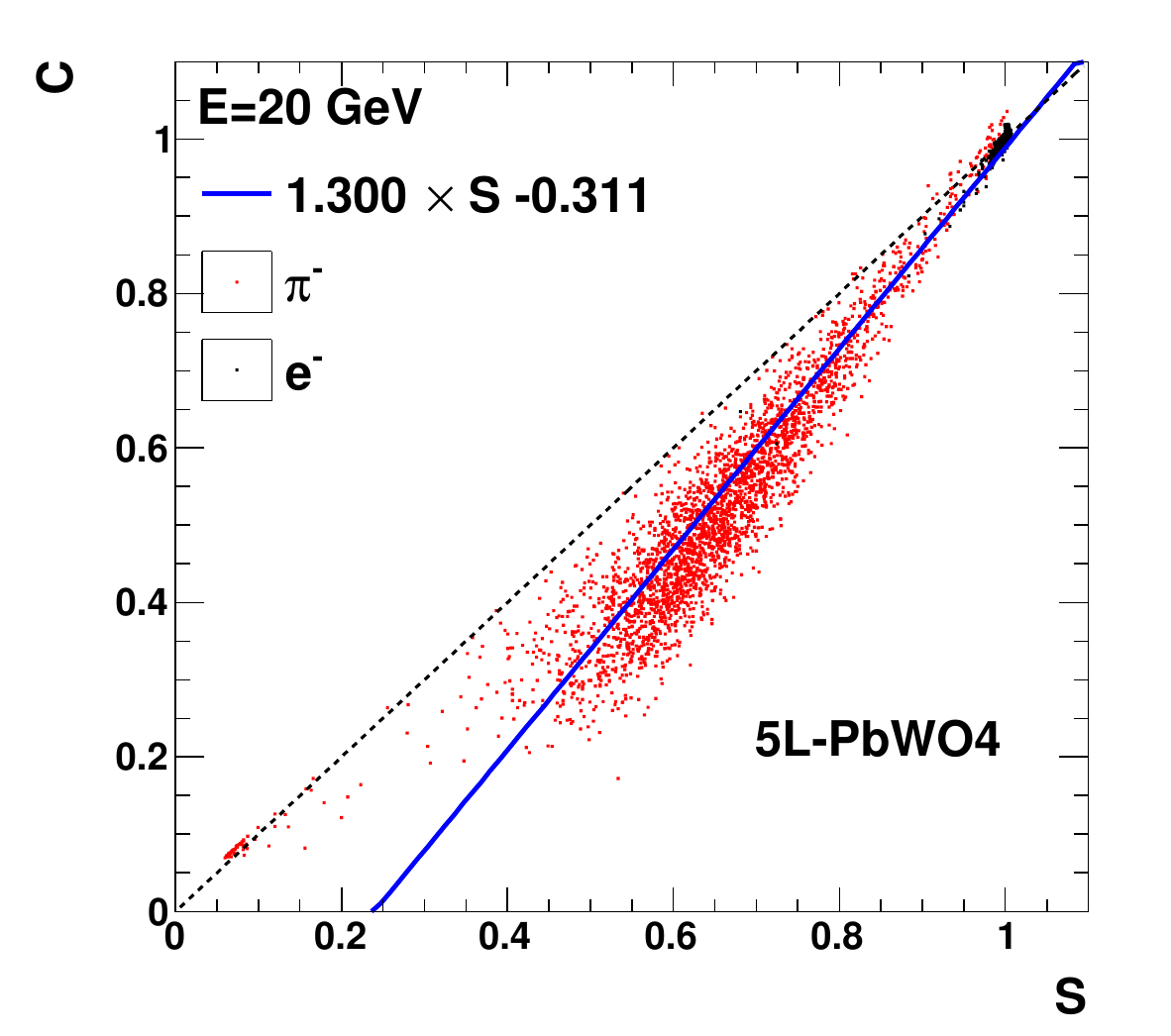}
  \end{center}
  \caption{
 Correlation between scintillation yield ($S$) and  Cherenkov yield ($C$), using a calibration based on 20\,GeV electrons, 
 for the homogeneous PbWO4 hadronic calorimeter (5L-PbWO4).
  The blue line shows the $\chi^2$ fit of the pion data.
  }
  \label{fig:correlation_pbwo4}
\end{figure}

Simulations have been performed for a single tower made from  PbWO4 crystals.
It has 5 blocks, each of which has the size 20x20x20\,cm. This configuration, to be
called  5L-PbWO4, corresponds approximately to 5 $\lambda_I$, which should give about 98\% containment for 1\,GeV pions.
Scintillation and Cherenkov lights are collected from the same crystals. 
As with the sampling calorimeter, 3000 events have been simulated for each particle type.
The estimated relative responses are $(h/e)_S = 0.58$ and $(h/e)_C=0.39$ for a 20\,GeV pion beam, corresponding to $\kappa = 0.67$. 
Fig.~\ref{fig:correlation_pbwo4} shows the correlation
between the $S$ and $C$ photon counts for $20$\,GeV pions after calibrating the signal to electron beams.
It can be seen that the spread of the signal is significantly smaller than for the sampling 40L-PFQ calorimeter (see Fig.~\ref{fig:correlation}).

\begin{figure}[hbt!]
\begin{center}

   \subfloat[5 GeV] {
   \includegraphics[width=0.48\textwidth]{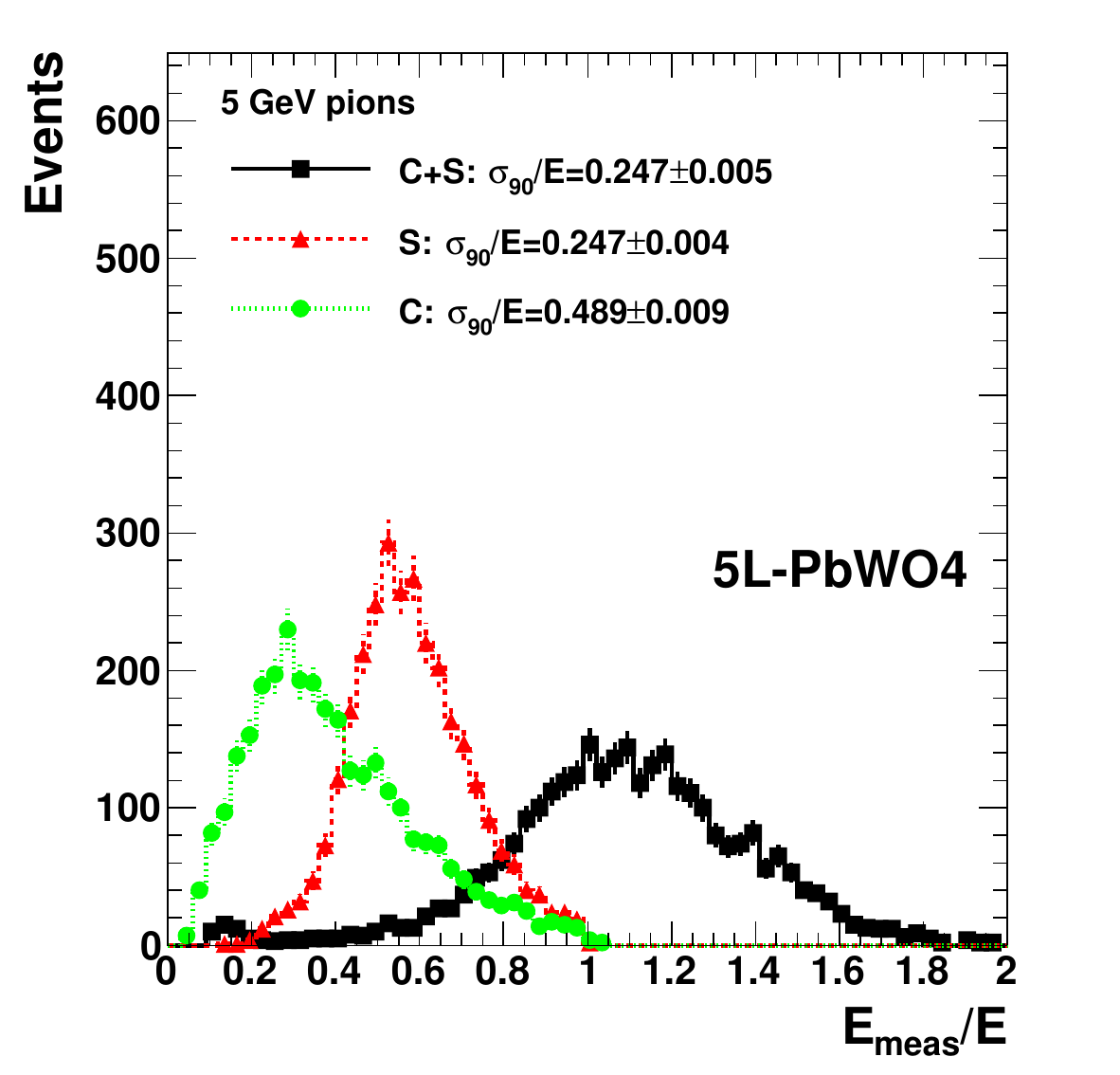}\hfill
   }
   \subfloat[10 GeV] {
   \includegraphics[width=0.48\textwidth]{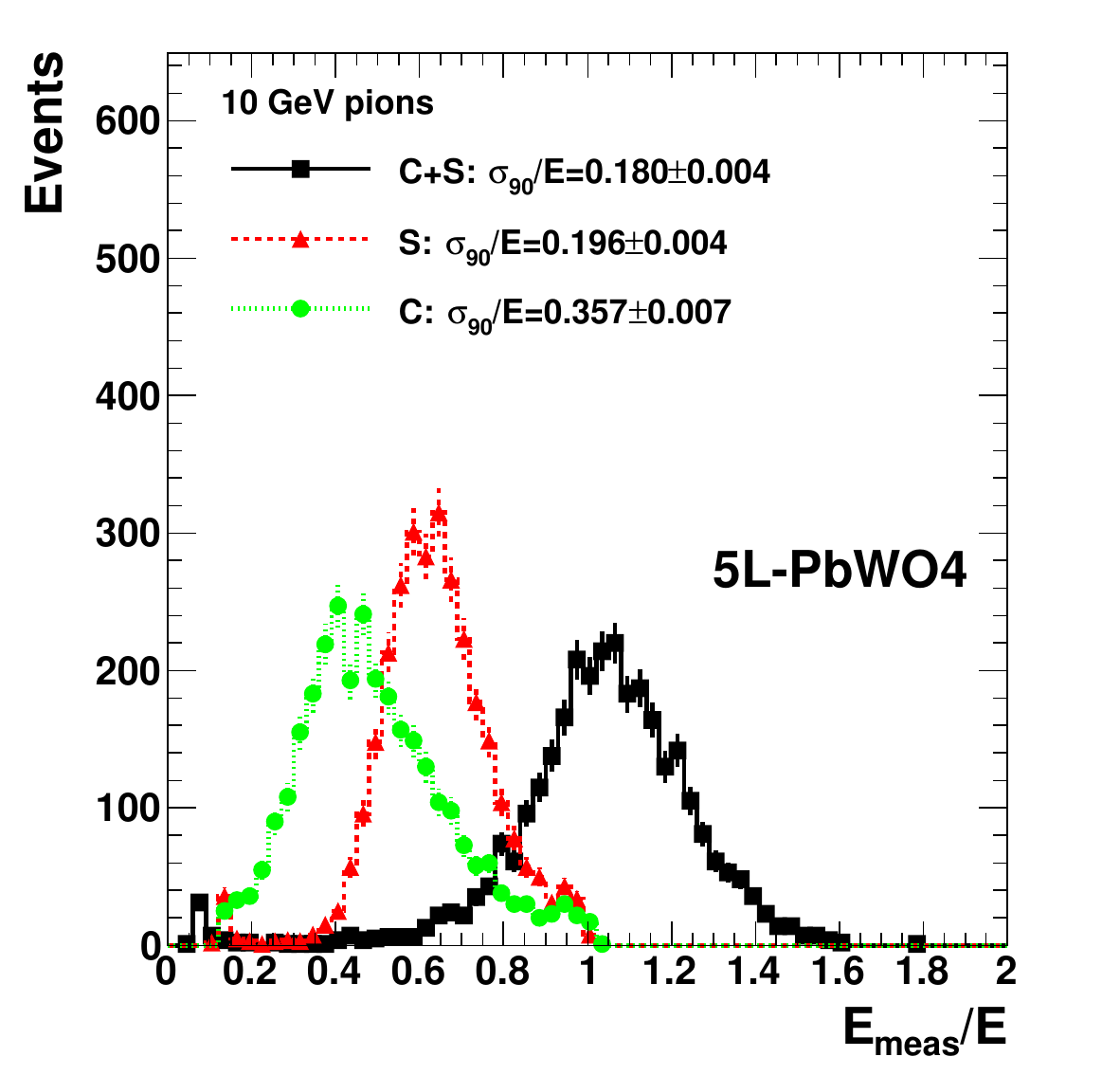}
   }

   \subfloat[20 GeV] {
   \includegraphics[width=0.48\textwidth]{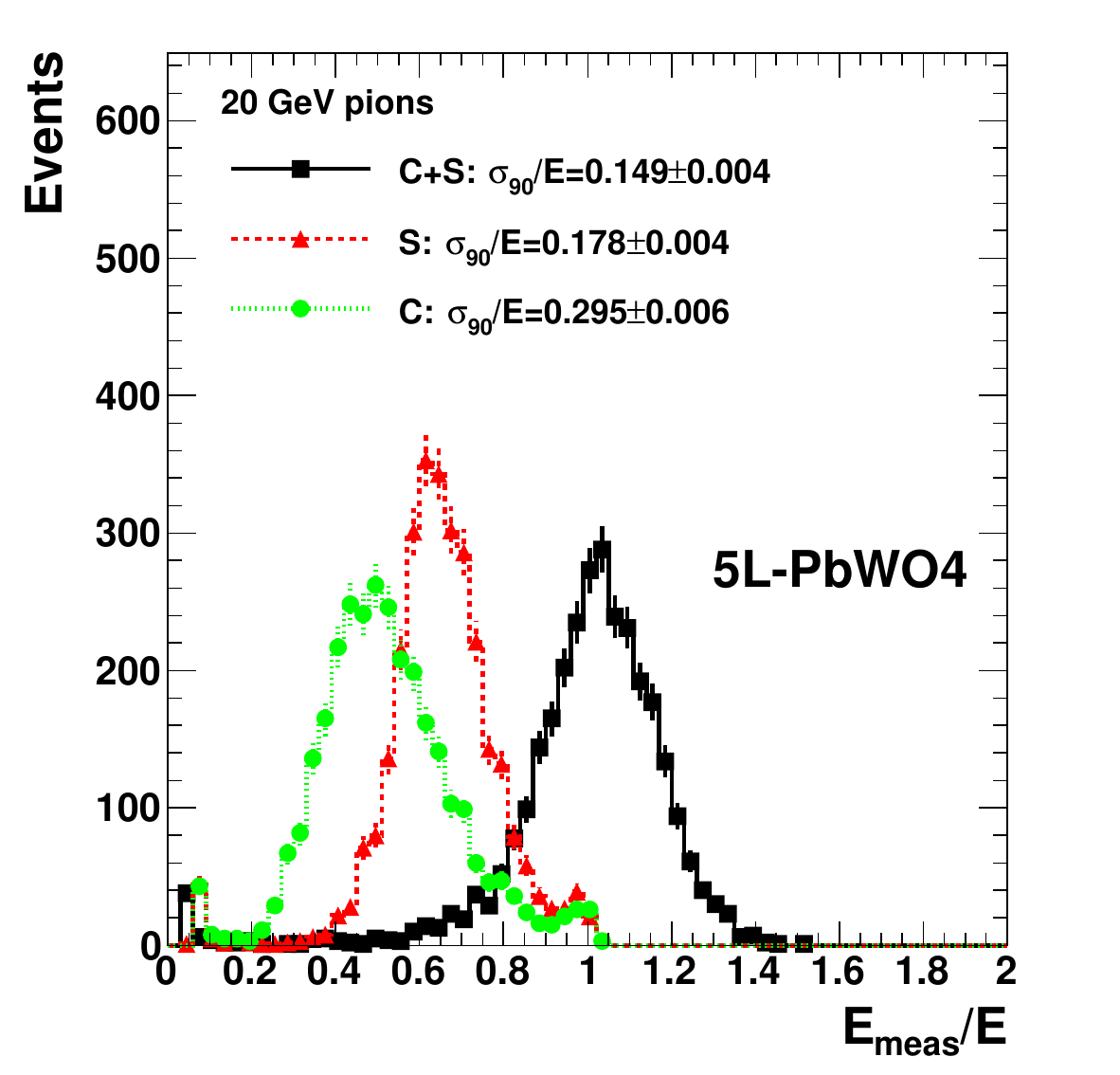}\hfill
   }
   \subfloat[40 GeV] {
   \includegraphics[width=0.48\textwidth]{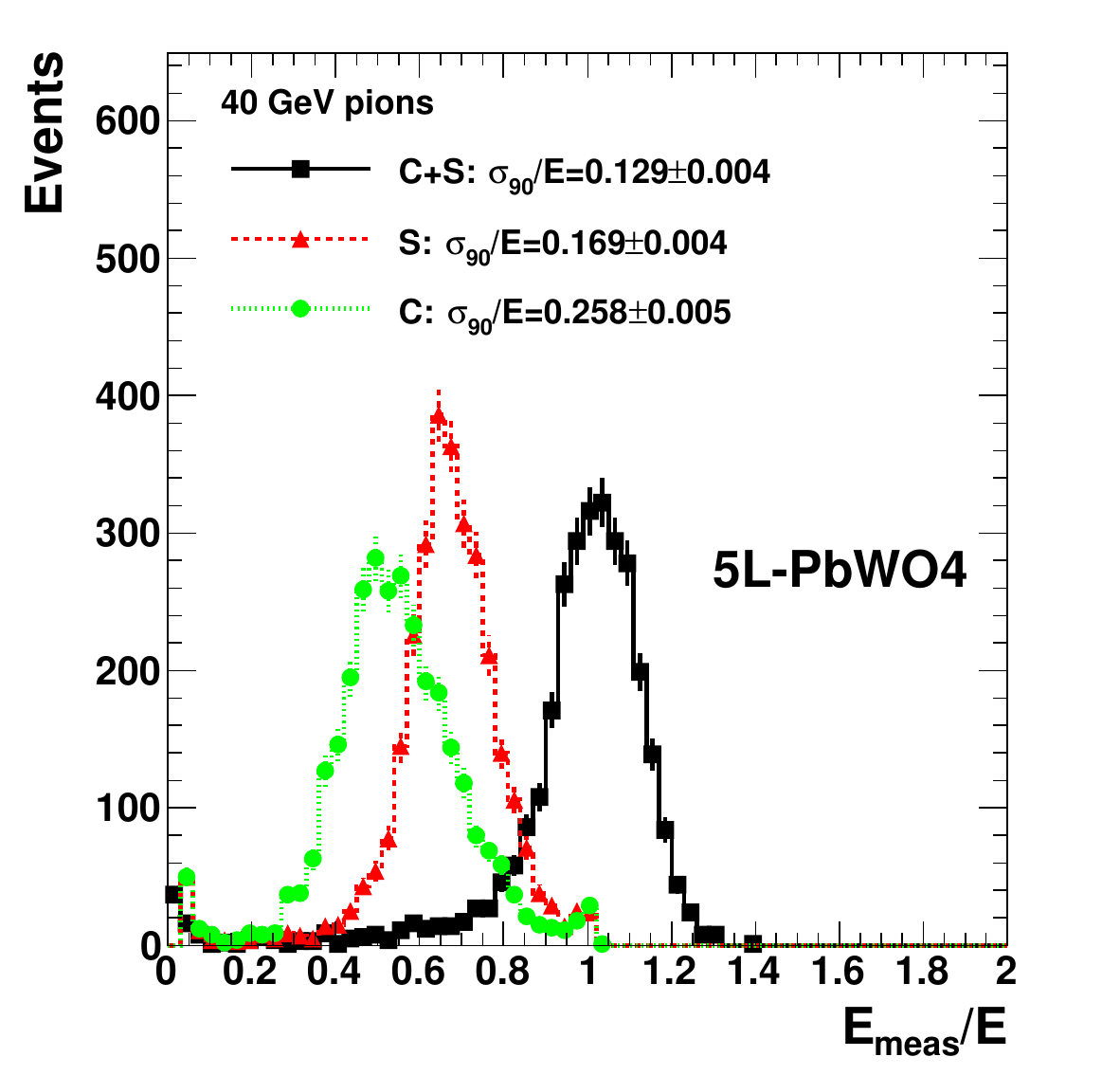}
   }
\end{center}
\caption{
The distributions of the $S$, $C$ and $S+C$ photons for  pion  beams with 
energies between 5 -- 40\,GeV for the PbWO4 crystal-based calorimeter. 
The mean and the $\sigma_{90}$ of the signal profiles are also shown.}
\label{fig:corrections2pbwo4}
\end{figure}

Figure~\ref{fig:corrections2pbwo4} illustrates the resolutions of $S$ and $C$ lights for the  5L-PbWO4 
hadronic calorimeter for  pion beams from 1 to 20\,GeV. After using Eq.~\ref{eqene}, the resulting
distribution for $S+C$ lights are shown with the black lines. After the correction, 
the resolution shows an improvement by 26\% for a 20\,GeV beam, 
compared to the $S$ light alone, with the improvement being smaller for lower energies. This demonstrates 
that the homogeneous calorimeter can benefit from the dual readout approach. 
In addition, the overall resolution of the $S$ signal  is better than for the sampling calorimeter with polystyrene active material.

\subsection{40 layer sampling  calorimeter without passive absorber (40L-PQ)}
\label{sec:40PQ}
An alternative design of the sampling calorimeter was checked. It uses the same geometry as that discussed
in Sect.~\ref{sampling} but removing the iron plates of  width 1.8\,cm, and replacing them
with  4\,cm quartz plates, which will act as absorber  and as the active layer for Cherenkov light.
The original 0.5\,cm quartz layer was removed.
The length of the calorimeter had to be increased to achieve about 5 $\lambda_I$, thus the 
overall design will not fit in the envelop of the CLD detector.  
The simulation of one tower of such detector was performed.
However, no improvements in the resolution after using the $C$ photons was observed.
The results were found to be similar to Figs.~\ref{fig:correlation} and ~\ref{fig:corrections1}

It is likely that such designs do not address the main physics goal of measuring $S$ and $C$ lights from the same particles produced in the shower, a feature which can easily be achieved by homogeneous crystal detectors. 

\subsection{250 layer sampling without passive absorber (250L-PQ)}
\label{sect:250L}

Another design, 250L-PQ ("polystyrene-quartz"), 
is similar to that discussed in Sect.~\ref{sec:40PQ}.
The number of layers is changed from 40 to 250. Each layer had 0.5~cm polystyrene and
0.5~cm quartz, separated by a foil of the width 0.5~mm. 
The size of the tower was increased appropriately, both in the transverse ($X-Y$) and the longitudinal ($Z$) direction.
The transverse size was set to 50\,cm (i.e. approximately one $\lambda_I$).
The longitudinal size was 313\,cm, which is too large for inclusion in  a realistic detector design.

The calculated parameters for this design are $\kappa=0.73$, $(h/e)_S=0.65$ and $(h/e)_C=0.52$ for 20\,GeV energies. This design shows some improvement after including the $C$ light detected from the quartz layers. 
The improvement was about 13\% for 20\,GeV beam, but it was smaller
for lower energies. This improvement will be discussed later.

\subsection{200 layer calorimeter with passive absorber (200L-PFQ)}
\label{sec:200L}

Finally, a calorimeter with passive layers was tested. This tower had 200 layers. Each layer had a passive layer of steel (0.3\,cm), and two active layers of polystyrene and quartz, each with the size of 0.3\,cm. The transverse size of this model was adjusted to  36\,cm, which corresponds to about one interaction length.
The calculated parameters for this design are $\kappa=0.52$, $(h/e)_S=0.77$ and $(h/e)_C=0.56$ for
20\,GeV energies.

This design, to be called 200L-PFQ,  
also shows some improvements in the resolution after including $C$ photons.
The improvement is about 5\% for a 20\,GeV beam, i.e. is somewhat smaller than for the 250-layer design discussed before,
and significantly smaller for the 5L-PbWO4 homogeneous crystal calorimeter.

\section{Summary of the simulation studies}

This section summarizes the \geant\ simulations for the variations of the single HCAL towers for the design 
choices considered in the previous section.

\begin{figure}[hbt!]
\begin{center}

   \subfloat[40L-PFQ] {
   \includegraphics[width=0.48\textwidth]{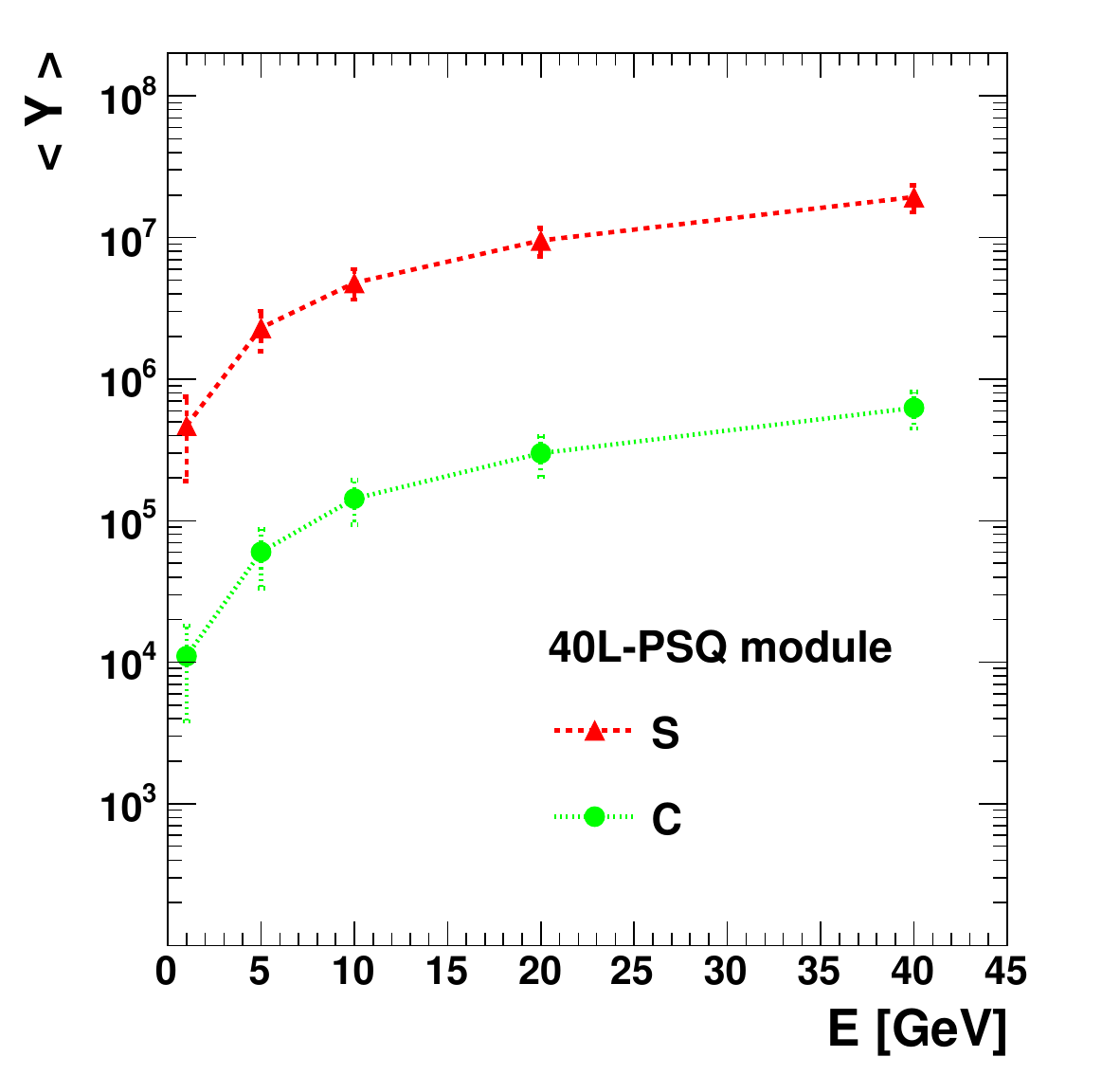}\hfill
   }
   \subfloat[5L-PbWO4] {
   \includegraphics[width=0.48\textwidth]{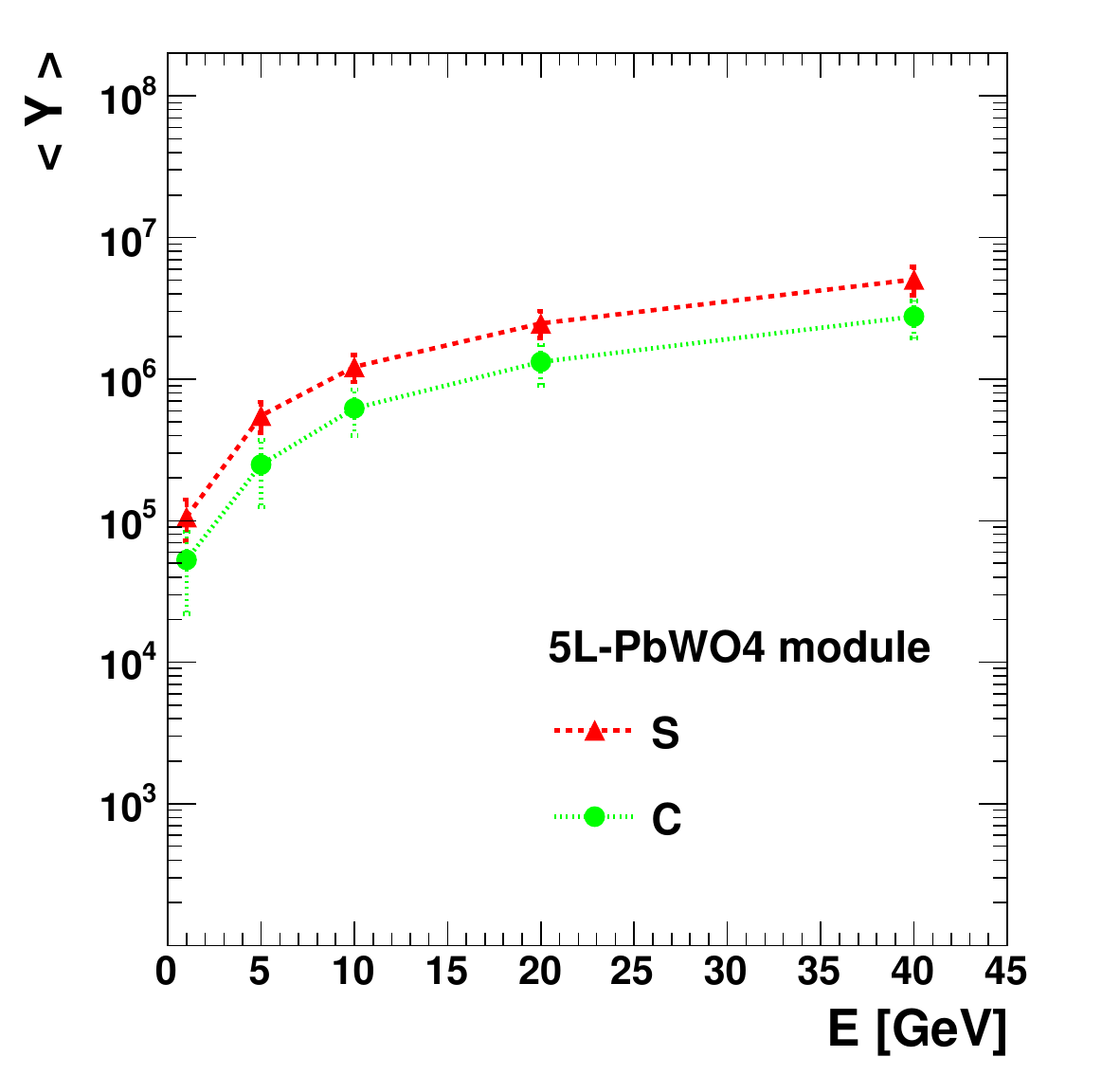}
   }

   \subfloat[250L-PQ] {
   \includegraphics[width=0.48\textwidth]{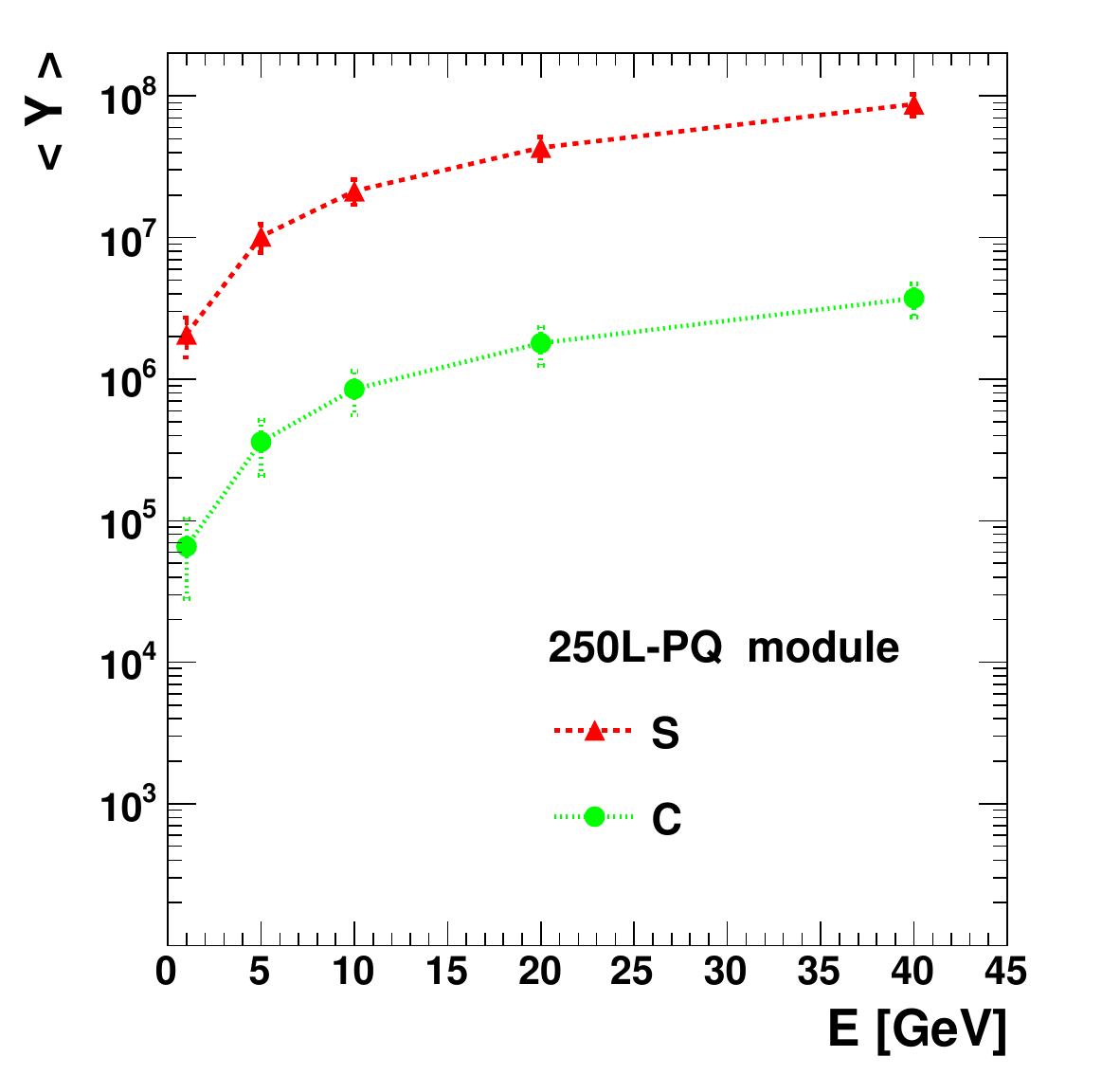}\hfill
   }
   \subfloat[200L-PFQ] {
   \includegraphics[width=0.48\textwidth]{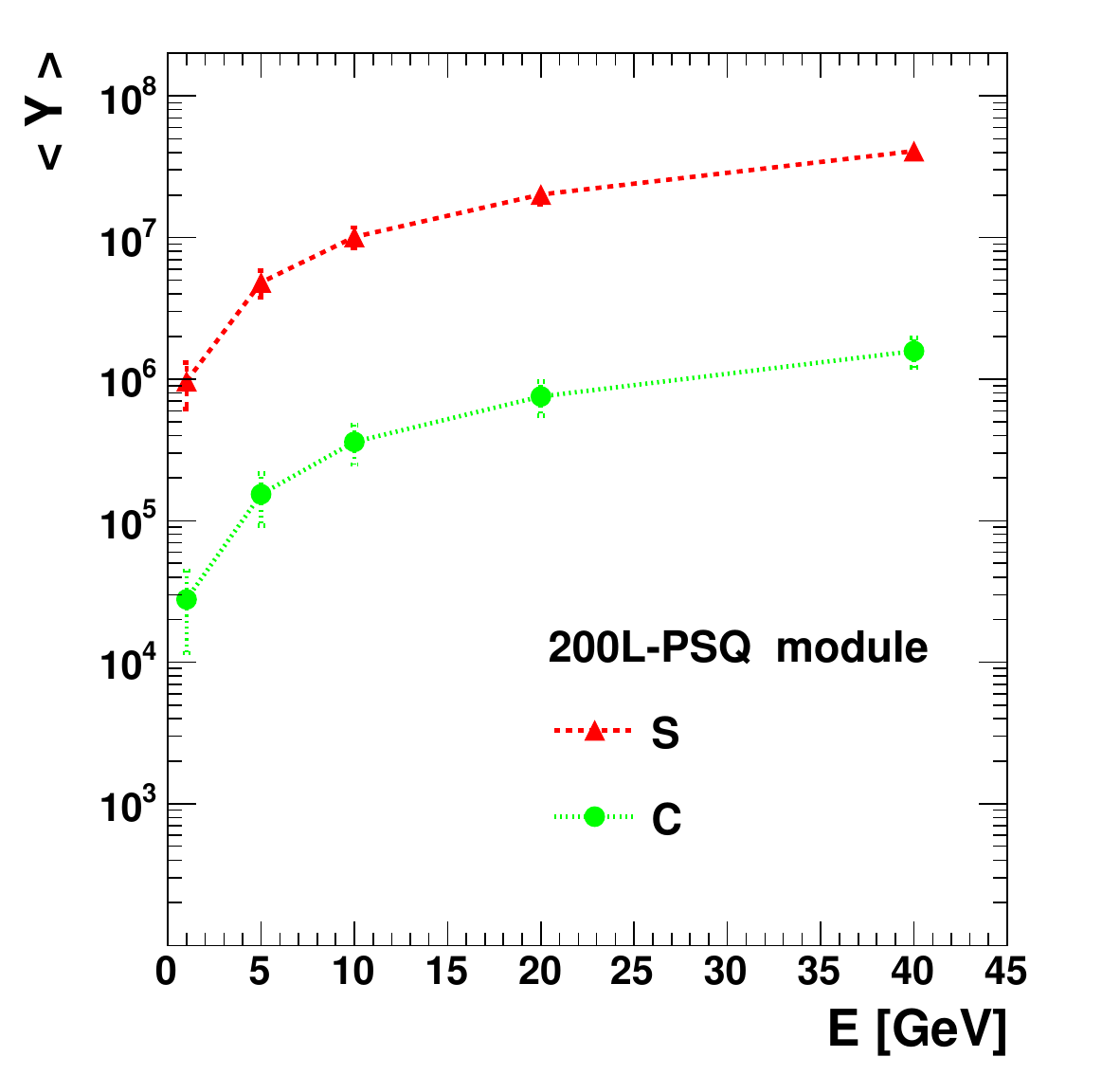}
   }
\end{center}
\caption{
Average number of optical photons for scintillation ($S$), Cherenkov ($C$) lights for 1 -- 40~GeV pions
for 4 different styles of the calorimeters (see the text). The error bars show the RMS for each energy.
(a) The standard calorimeter with 40 layers where each layer has 2 active layers (polystyrene and quartz) and the passive layer (Fe) of the with of 1.8\,cm; (b) A homogeneous calorimeter with 5 blocks of PbWO4 discussed in Sect.~\ref{crystal}; (c) A sampling calorimeter with 250 layers, where each layer is made from polystyrene and quartz, see Sect.~\ref{sect:250L}; 
(d) A sampling calorimeter with 200 layers, where each layer is made of polystyrene (active), quartz (active) and passive layer (steel), see Sect.~\ref{sec:200L}.
}
\label{fig:yields}
\end{figure}

Fig.~\ref{fig:yields} shows the average number of simulated optional photons as a function of the beam energy.
The vertical error bars indicate the RMS for each energy. The plots demonstrate  significant photon yields, 
ranging from $10^4$ to $10^8$, before applying any corrections. As expected, the average 
rate of the scintillation photons is larger than the rate of the Cherenkov photons. In the case of the PbWO4
tower, the rate of the $S$ and $C$ photons is closest among the other technology choices.

The demonstrated yields  of the $S$  and $C$ photons  with the single photon precision
are CPU intensive, and are impractical for full simulations of even a single calorimeter tower. 
For example, a simulation of a single collision with 40\,GeV 
pions, leading to $5\times10^6$ photons,   takes about 20\,min per event for the 5L-PbWO4 style calorimeter using a
Xeon(R) CPU E5-2650 v4 @ 2.20GHz. The simulation time increases to 40\,min when using the 250 layer calorimeters that
produce up to $10^8$ photons for the 4\,GeV beams. For the $4\lambda_I$ towers, the time further increases by a factor 3.
Therefore, alternative methods need to be investigated for a proper inclusion of such effects in the simulations of complete detectors.

\begin{figure}[hbt!]
\begin{center}

   \subfloat[40L-PFQ] {
   \includegraphics[width=0.48\textwidth]{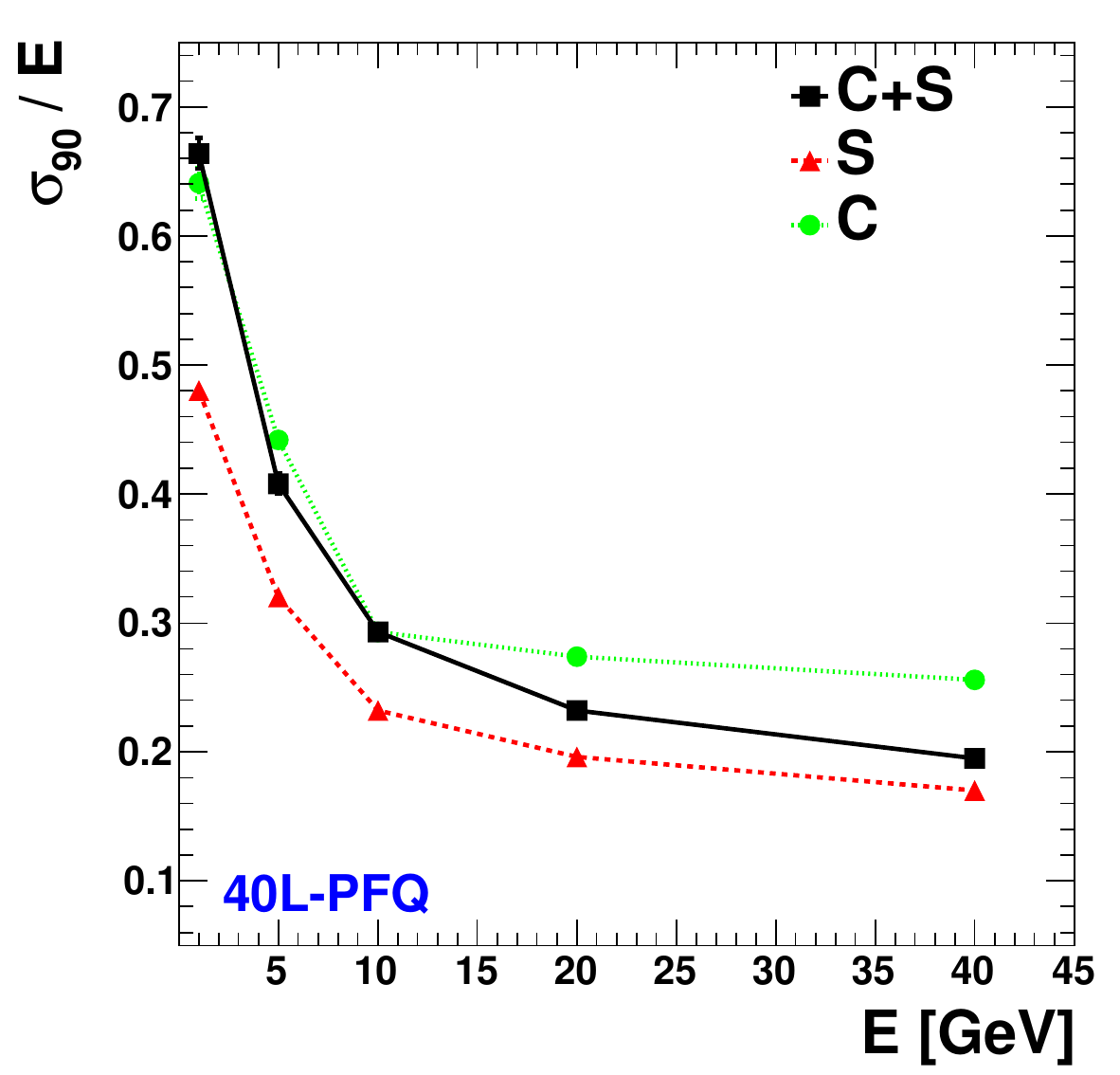}\hfill
   }
   \subfloat[5L-PbWO4] {
   \includegraphics[width=0.48\textwidth]{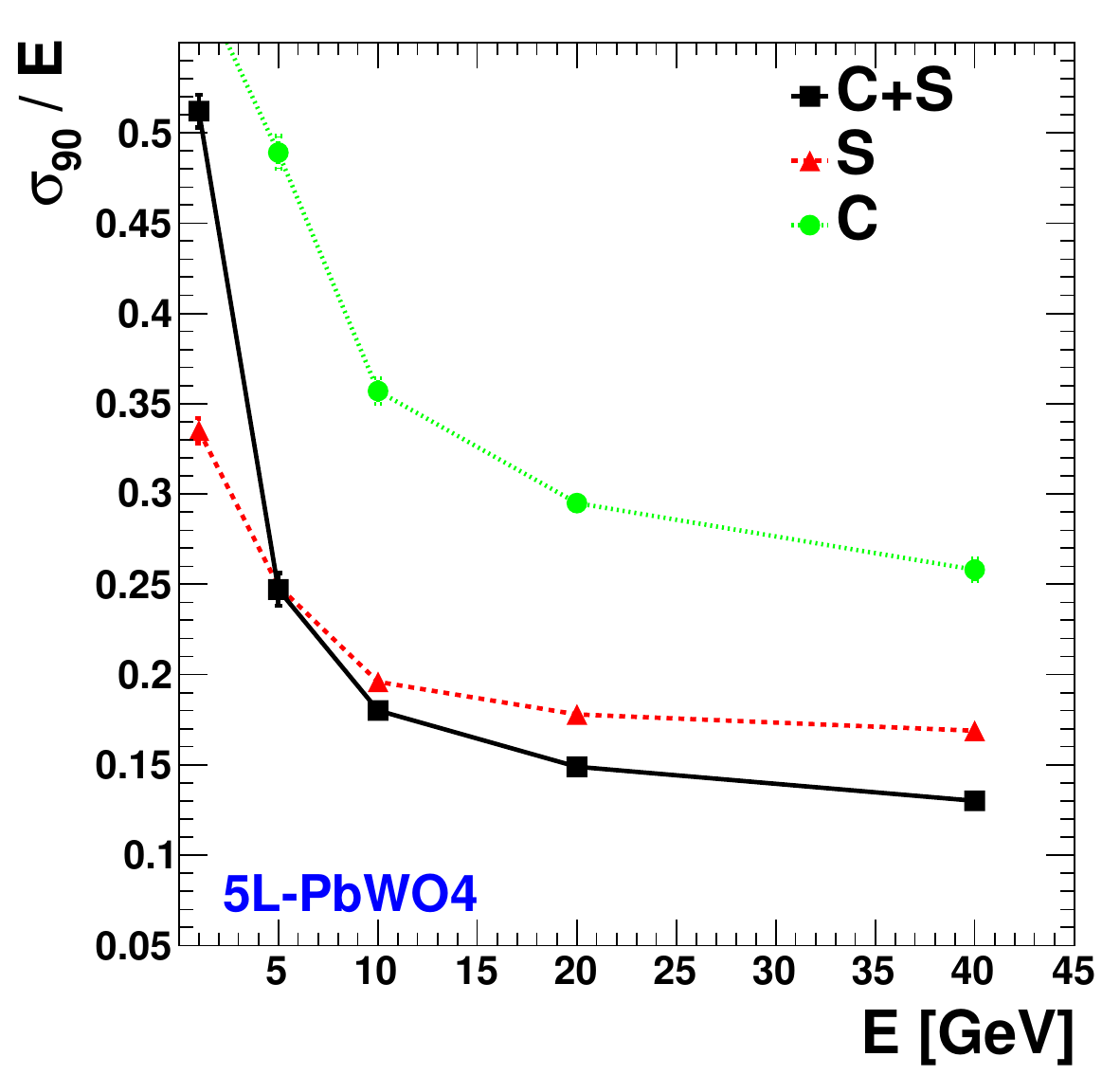}
   }

   \subfloat[250L-PQ] {
   \includegraphics[width=0.48\textwidth]{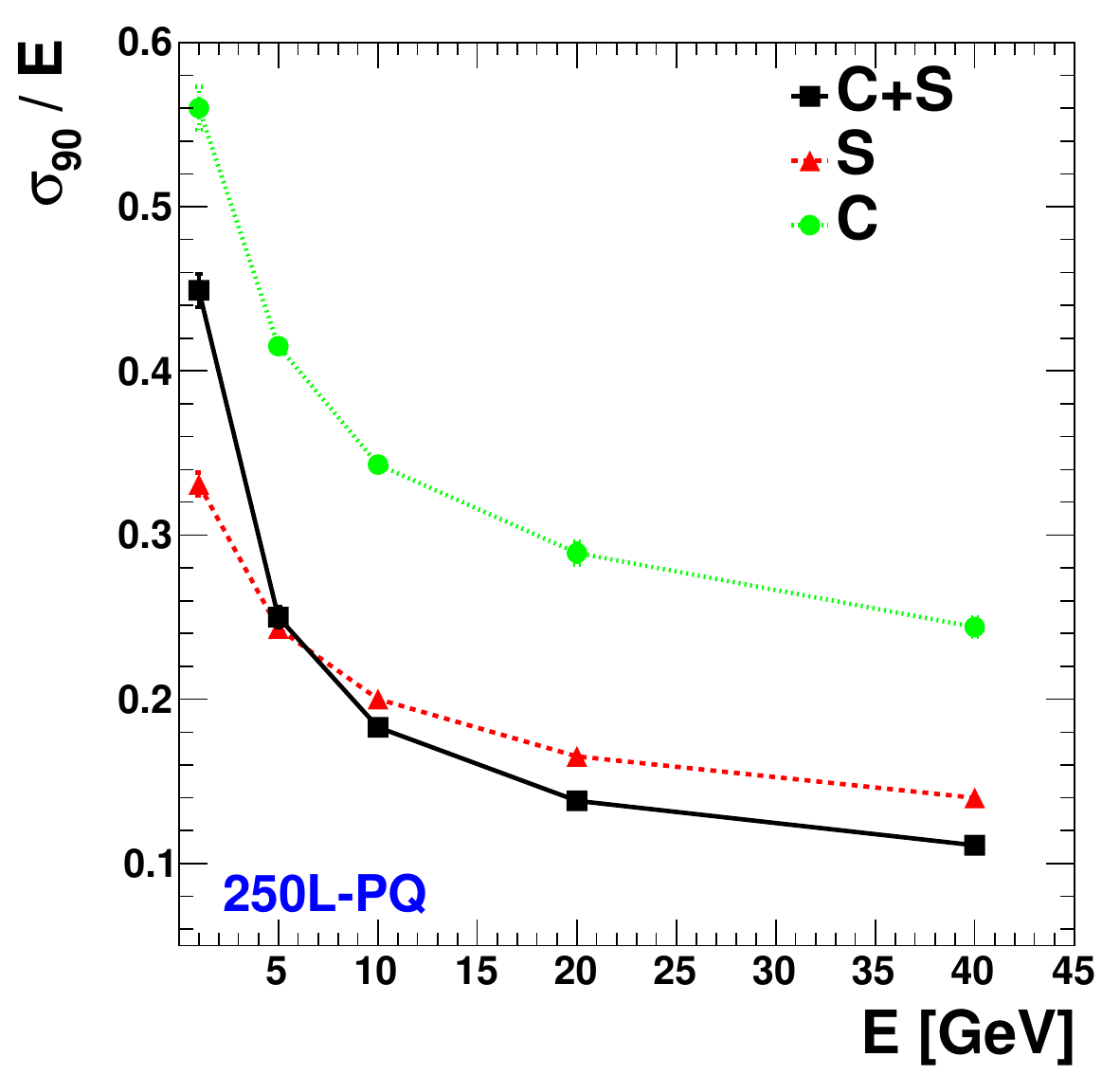}\hfill
   }
   \subfloat[200L-PFQ] {
   \includegraphics[width=0.48\textwidth]{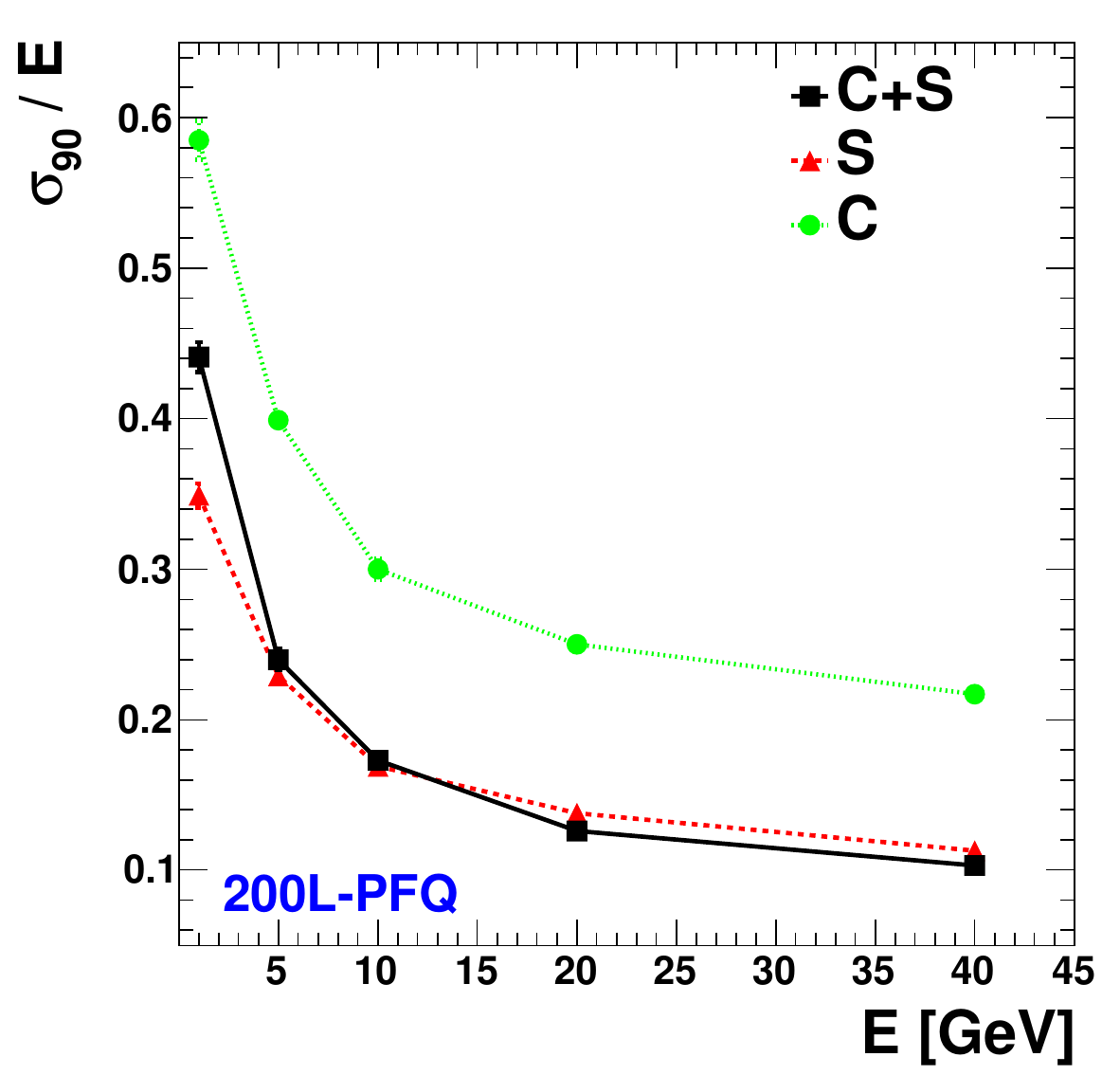}
   }
\end{center}
\caption{
Resolution of $S$ , $C$ and $S+C$  photons for 1 -- 40~GeV pions
for 4 different styles of the calorimeters (see the text).
(a) The standard calorimeter with 40 layers where each layer has 2 active layers (polystyrene and quartz) and the passive layer (Fe) of the with of 1.8\,cm; (b) A homogeneous calorimeter with 5 blocks of PbWO4 discussed in Sect.~\ref{crystal}; (c) A sampling calorimeter with 250 layers, where each layer is made from polystyrene and quartz, see Sect.~\ref{sect:250L}; (d) 
A sampling calorimeter with 200 layers, where each layer is made of polystyrene (active), quartz (active) and passive layer (steel), see Sect.~\ref{sec:200L}.
}
\label{fig:summary}
\end{figure}

Figure~\ref{fig:summary} shows the summary of resolution studies for the calorimeters considered in the previous section. The largest effect from the inclusion of Cherenkov signal is seen for the homogeneous crystal calorimeter. Some improvements in the resolution are seen for high-granularity 
calorimeters with 200 and 250 layers. This
improvement is only seen for $>10$\,GeV poins, while the inclusion of the $C$ light in the reconstruction makes 
the resolution worse at lower energy.
We expect additional segmentation in the transverse direction may help take advantage of $C$ photon counts for dual readout corrections.
The overall improvements of the $S+C$ photons at 40\,GeV, with respect to the $S$, is about 30\% for the 5L-PbWO4 tower, 26\% for the 250L-PQ and 10\% for 200L-PFQ. For the polystyrene-quartz designs, it is expected that the improvement will get larger as the number of longitudinal layers is increased .

\begin{figure}[hbt!]
\begin{center}

   \subfloat[40L-PFQ (4$\lambda_I$ tower width)] {
   \includegraphics[width=0.48\textwidth]{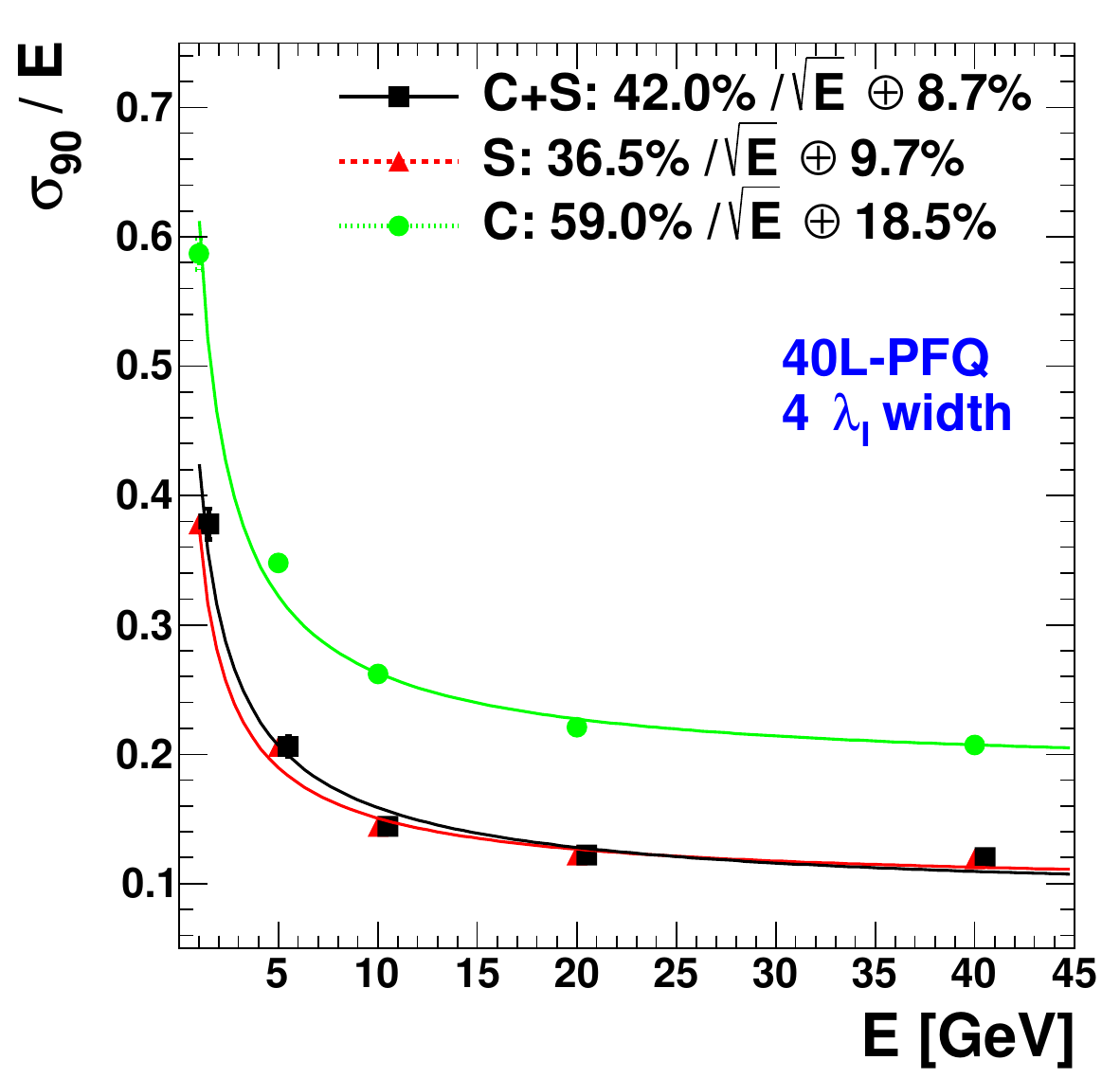}\hfill
   }
   \subfloat[5L-PbWO4 (4$\lambda_I$ tower width)] {
   \includegraphics[width=0.48\textwidth]{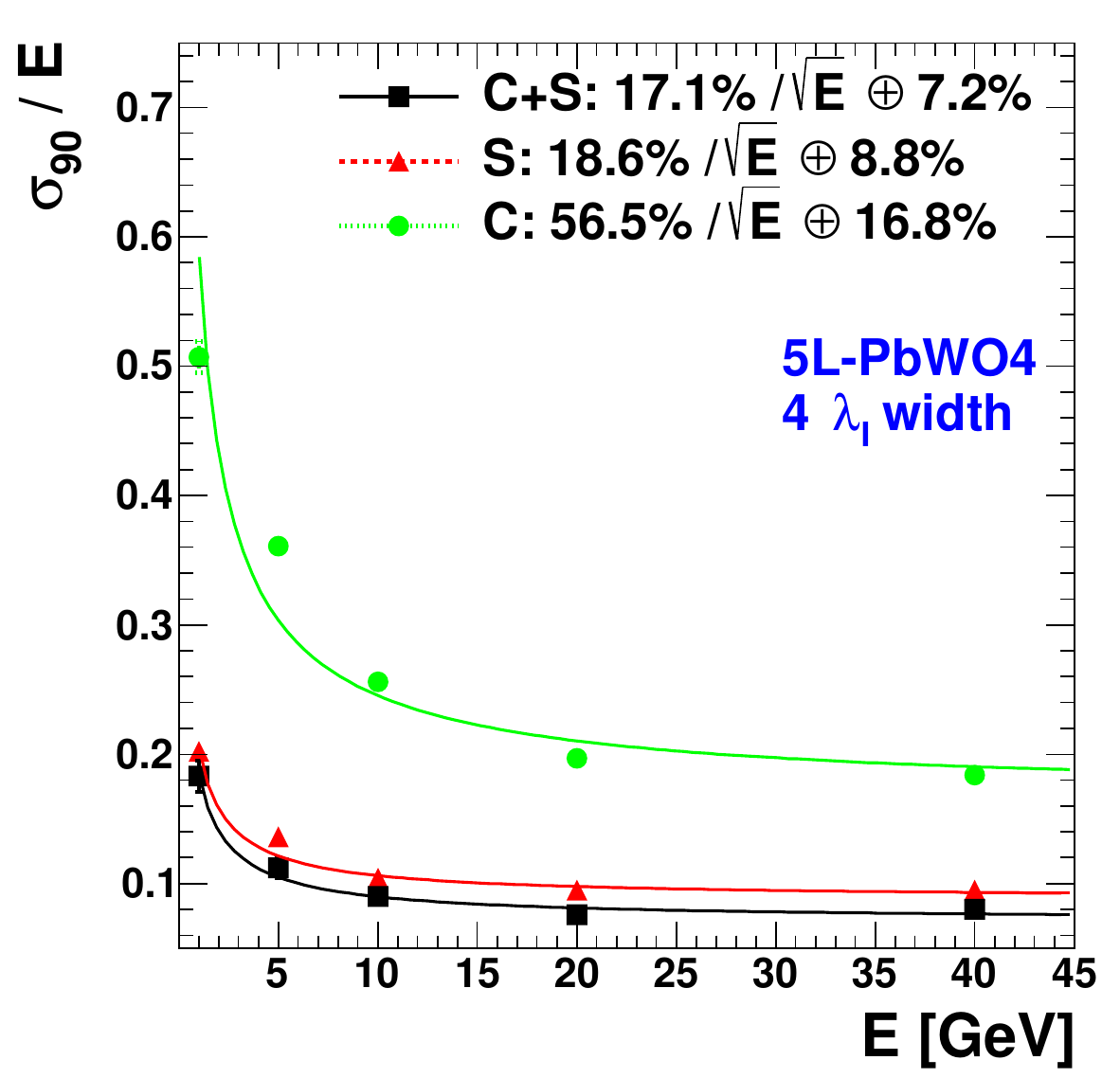}
   }
\end{center}
\caption{
Resolution of $S$,  $C$ and  $S+C$ photons for 1 -- 40~GeV pions
for the 40L-PFQ and 5L-PbWO4 towers with the $4\lambda_I$ width.
The lines show the fit results with Eq.(\ref{res}).
These figures should be compared the $1\lambda_I$ designs shown in Fig.~\ref{fig:summary}(a)(b)
}
\label{fig:summary4lambda}
\end{figure}

As discussed above, to limit the CPU usage, the above studies were performed for the towers with $1\lambda_I$ width. This leads to a non-negligible shower leakage, deteriorating the resolution. 
The simulations were performed 
for the 40L-PFQ and 5L-PbWO4 towers 
with a $4\lambda_I$ lateral width.
The resolution for the $S$ signals in the  40L-PFQ
tower improves by  40\% compared to the $1\lambda_I$ width design since the  shower leakage was minimized. 
This 40L-PFQ design indicates full 
compensation for the scintillation photons, with  $(h/e)_S \simeq  1$.
Thus, as expected, adding $C$ photons  to the $S$ signal  
did not lead to improvements in the resolution 
over the scintillation light alone. 
Figure~\ref{fig:summary4lambda}(a) compares the $S$ and $C$ resolutions. 

It worth mentioning
that the number of $C$  photons was about 3\% of the total number of optical photons. These photons were concentrated in the narrow cone around the incoming particles, thus the $S$ signal dominates the regions beyond  $0.5\lambda_I$ cone with the center at $X=Y=0$. 

As a check, the 200L-PQF design was extended to the $4\lambda_I$ size
in the transverse direction. This design did not show an obvious
improvement in the resolution after adding  $C$ photons to the $S$ signal.
This check was performed with a smaller number of events (1000) as it required significant CPU resources.

For the  5L-PbWO4 tower with the $4\lambda_I$ width, the resolution of the $S$ photons was also improved by almost 50\% compared to the $1\lambda_I$ width tower, see Figure~\ref{fig:summary4lambda}(b). 
This tower with the reduced lateral shower leakage 
is  non-compensating, with  $\kappa=0.51$, $(h/e)_S=0.80$ and $(h/e)_C=0.61$ for
20\,GeV energies. Adding the $C$ signal to the $S$ photons improves the resolution by roughly 15\% at 10 GeV.

The resolution $\sigma_{90}/E$ of the $S+C$ photons for 5L-PbWO4 
with the $4\lambda_I$ tower can be described by a linear sum
\begin{equation}
\frac{\sigma_{90}} {E} = \frac{a}{\sqrt{E}} \oplus  b,
\label{res}
\end{equation}
where $a$ is a sampling (stochastic) term, which dominates the most of the relevant energy spectrum, and 
$b$ is a constant term attributed to high-energy part of the resolution.  
The stochastic term for $S+C$ photons was found to be 17.1\%
from the $\chi^2$ fit of  Fig.~\ref{fig:summary4lambda}(b), while
$a=$18.6\% for $S$ photons.
The stochastic term for the 40L-PFQ sampling tower with the reduced
lateral shower leakage (shown in Fig.~\ref{fig:summary4lambda}(a)) 
was found to be $37 (42)\%$  for $S$ (or $S+C$) photons.

The results of this paper can be compared to the studies of the RD52 lead-fiber dual-readout calorimeter presented in \cite{Lee:2017xss}.
In this comparison, one can use the $\sigma_{90}/E$ values shown  in Fig.~\ref{fig:summary}. To obtain the RMS values, which are typically used to define resolutions, one should multiply $\sigma_{90}/E$ by 1.25.
The resolution for 20~GeV pions for the RD52 hadronic calorimeter was 13.5\% (Fig.~39 of \cite{Lee:2017xss}). According to our simulations,  the $4\lambda_I$ 40L-PFQ tower
has $\sigma_{90}/E$=12.2\% (for $S$ or $S+C$ at 20~GeV), which roughly translates to the $\sigma/E=15.2\%$.
In the case of 5L-PbWO4, $\sigma_{90}/E$=9.6\% (for $S$) and  $\sigma_{90}/E$=7.6\% (for $S+C$),
which correspond to $\sigma/E$=12\% and 9.5\%, respectively. 
Thus, the simulation predicts that 5L-PbWO4 tower overperforms the RD52 HCAL, while the sampling calorimeter underperforms it.
The resolution for the sampling calorimeters with 200 and 250 layers is expected
to be very similar to the RD52 calorimeter. 
In all cases, the simulations indicate a smaller overall correction from the $C$ light
compared to the RD52 lead-fiber dual-readout calorimeter. 
The obtained $\sigma/E$=9.5\% resolution for 20~GeV pions for the 5L-PbWO4 tower is similar to the resolution of single pions in a dual-readout homogeneous calorimeter using the CaTS simulation framework \cite{Wenzel_2012}. 

The improvement in the relative resolution after adding the $C$ photons  depends on the width (and the mean) of this signal. On the pure statistical bases,
when the $C$ signal has the same width and the mean as the $S$ photons, the improvement in $\sigma_{90}/E$ for the $S+C$ signal is expected to be as large as 30\% compared to the
$S$ signal alone. However, the distribution for the $C$ photons is typically wider than for the $S$ signals. This reduces the potential gain that can be 
obtained from including the $C$ photons. Nevertheless, improvements in the relative  resolution are still quite noticeable for the 5L-PbWO4 tower.

\section{Conclusion}

In this paper we report the  results of the \geant\ simulation of optical photons
created as the results of collisions with single particles with the energy 1 -- 40~GeV.
The simulations were performed for  the traditional HCAL tower considered for the ILC, CLIC, CLD and similar experiments, but after adding additional  active layers to collect  \cere\ signals.  For this design,  the $S$  and $C$ photons can be collected separately from the active layers made from polystyrene and quartz. 
The simulations of optical photons were performed  with the single-photon precision.
The simulations did not demonstrate  improvements in resolution after including 
$C$ signals.  On the contrary,  we observe a larger smearing for $S+C$ photons  compared to the $S$ light alone when using the $1\lambda_I$ width towers.
If the tower width is increased to $4\lambda_I$ to minimise the lateral shower leakage, the resolutions of $S$ and $S+C$ distributions are the same as for a fully compensated sampling calorimeter.
Thus, the traditional sampling calorimeter design  with several dozed layers used for the ILC, CLIC and CLD detectors cannot benefit  from  the dual-readout energy correction.
The obtained $S$ (or $S+C$) photon resolution is slightly worse than for the RD52 lead-fiber dual-readout calorimeter \cite{Lee:2017xss}. 

When using the large number (i.e. $\geq 200$) of sampling layers for the sampling HCAL tower, the inclusion of  \cere\ light leads to a noticeable improvement in the resolution above 10\,GeV. For 40\,GeV pions, the resolution improvements vary between 10\% - 26\%, depending on the number of layers. 
With the large-layer design presented in this paper, the highly-segmented calorimeters will not fit into the nominal envelop  of the CLD detector, thus such towers cannot be considered in practical situations. Therefore, materials with a larger interaction length, such as tungsten (for the passive layer) or lead glass (for the active layer made of quartz) should be considered. 

A HCAL tower constructed from PbWO4 crystals shows the best resolution for scintillation photons compared to the studied sampling towers. In addition, the \geant\; simulation demonstrates a further  improvement in the resolution of optical photons after adding Cherenkov photons. This improvement can be as large as 30\% for 40\,GeV hadrons,  but it is somewhat smaller at lower energies. Among the towers studied in this paper, 
this homogeneous calorimeter shows the largest improvement in resolution after including  Cherenkov photons, which is expected for calorimeters with a smaller $h/e$ than for the sampling calorimeter.
The \geant\, simulations indicate that the dual-readout correction to the resolution is smaller than for the RD52 lead-fiber dual-readout calorimeter \cite{Lee:2017xss}, but the overall $S+C$ resolution of the crystal calorimeter is better than for the RD52 calorimeter.  

To this date, this is a first attempt to simulate optical photons produced in 
a highly-segmented sandwich calorimeter from the first physics principles of the \geant\; program without simplifications to reduce the CPU usage. 
The most common technique in the past was to simulate energy deposits, rather than optical photons.
Previously, dual-readout Monte Carlo simulations have been performed for the RD52 fiber calorimeter \cite{Akchurin:2014xpa} and homogeneous calorimeters \cite{Magill_2012,Wenzel_2012,Lucchini:2020bac}.
Due to the significant CPU usage, the simulations  discussed in this paper are unpractical for the detector geometries beyond single towers. Therefore, alternative methods need to be investigated for the inclusion of dual readout in the simulations of complete HCAL geometries and collision events. For example, photons can be grouped together or the optical yields can be artificially reduced while preserving the balance between the Cherenkov and scintillation photon rates.  In addition, exascale supercomputers and the usage of GPU should enable such complex simulations.
                        
\section*{Data Availability}
The simulated data used in this study can be accessed via HepSim ~\cite{Chekanov:2014fga}.
The code used in this paper is available from \cite{githubc}.

\section*{Acknowledgments}
We thank M.~Lucchini for the discussion.
We gratefully acknowledge the computing resources provided by the Laboratory
Computing Resource Center at Argonne National Laboratory.
The submitted manuscript has been created by UChicago Argonne, LLC, Operator of Argonne National Laboratory (“Argonne”). Argonne, a U.S. 
Department of Energy Office of Science laboratory, is operated under Contract 
No. DE-AC02-06CH11357.  
Eno is supported via  U.S. Department of Energy Grant DE-SC0022045.

% example
% https://iopscience.iop.org/article/10.1088/1748-0221/15/11/P11005/pdf

%\clearpage
%\appendix
% some useful text to describe the layout https://flc.desy.de/hcal/basics/calorimetry/index_eng.html
%\section{A}

\bibliographystyle{JHEP}
\bibliography{references}

\providecommand{\href}[2]{#2}\begingroup\raggedright\begin{thebibliography}{10}

\bibitem{Sefkow:2015hna}
F.~Sefkow, A.~White, K.~Kawagoe, R.~P\"oschl and J.~Repond, \emph{{Experimental
  Tests of Particle Flow Calorimetry}},
  \href{https://doi.org/10.1103/RevModPhys.88.015003}{\emph{Rev. Mod. Phys.}
  {\bfseries 88} (2016) 015003}
  [\href{https://arxiv.org/abs/1507.05893}{{\ttfamily 1507.05893}}].

\bibitem{Magill_2012}
S.~Magill, \emph{Use of particle flow algorithms in a dual readout crystal
  calorimeter},
  \href{https://doi.org/10.1088/1742-6596/404/1/012048}{\emph{Journal of
  Physics: Conference Series} {\bfseries 404} (2012) 012048}.

\bibitem{AKCHURIN2005537}
N.~Akchurin, K.~Carrell, J.~Hauptman, H.~Kim, H.~Paar, A.~Penzo et~al.,
  \emph{Hadron and jet detection with a dual-readout calorimeter},
  \href{https://doi.org/10.1016/j.nima.2004.07.285}{\emph{Nuclear Instruments
  and Methods in Physics Research Section A: Accelerators, Spectrometers,
  Detectors and Associated Equipment} {\bfseries 537} (2005) 537–561}.

\bibitem{Lee:2017xss}
S.~Lee, M.~Livan and R.~Wigmans, \emph{{Dual-Readout Calorimetry}},
  \href{https://doi.org/10.1103/RevModPhys.90.025002}{\emph{Rev. Mod. Phys.}
  {\bfseries 90} (2018) 025002}
  [\href{https://arxiv.org/abs/1712.05494}{{\ttfamily 1712.05494}}].

\bibitem{Pezzotti:2022ndj}
I.~Pezzotti et~al., \emph{{Dual-Readout Calorimetry for Future Experiments
  Probing Fundamental Physics}},
  \href{https://arxiv.org/abs/2203.04312}{{\ttfamily 2203.04312}}.

\bibitem{Lucchini:2020bac}
M.T.~Lucchini, W.~Chung, S.C.~Eno, Y.~Lai, L.~Lucchini, M.-T.~Nguyen et~al.,
  \emph{{New perspectives on segmented crystal calorimeters for future
  colliders}},
  \href{https://doi.org/10.1088/1748-0221/15/11/P11005}{\emph{JINST} {\bfseries
  15} (2020) P11005} [\href{https://arxiv.org/abs/2008.00338}{{\ttfamily
  2008.00338}}].

\bibitem{Takeshita:2023fap}
T.~Takeshita and R.~Terada, \emph{{Simulation results of a New type of sandwich
  calorimeter, Double readout Sandwich Calorimeter (DSC) performance}},
  \href{https://arxiv.org/abs/2306.16325}{{\ttfamily 2306.16325}}.

\bibitem{Antonello:2018sna}
M.~Antonello et~al., \emph{{Tests of a dual-readout fiber calorimeter with SiPM
  light sensors}},
  \href{https://doi.org/10.1016/j.nima.2018.05.016}{\emph{Nucl. Instrum. Meth.
  A} {\bfseries 899} (2018) 52}
  [\href{https://arxiv.org/abs/1805.03251}{{\ttfamily 1805.03251}}].

\bibitem{CALICE:2012ami}
{\scshape CALICE} collaboration, \emph{{Calorimetry for Lepton Collider
  Experiments - CALICE results and activities}},
  \href{https://arxiv.org/abs/1212.5127}{{\ttfamily 1212.5127}}.

\bibitem{articleTESLA}
G.~Alexander, F.~Badaud, M.~Battaglia, T.~Behnke, M.~Berggren and
  S.~Bertolucci, \emph{Tesla technical design report, part iv: A detector for
  tesla}, {\emph{DESY-01-011, DESY-2001-011, DESY-01-011D, DESY-2001-011D,
  DESY-TESLA-2001-23, DESY-TESLA-FEL-2001-05, ECFA-2001-209} (2001) }.

\bibitem{SCHUWALOW2009258}
S.~Schuwalow, \emph{Calorimetry at the ilc detectors},
  \href{https://doi.org/https://doi.org/10.1016/j.nima.2008.08.122}{\emph{Nuclear
  Instruments and Methods in Physics Research Section A: Accelerators,
  Spectrometers, Detectors and Associated Equipment} {\bfseries 598} (2009)
  258}.

\bibitem{Bedeschi:2021nln}
F.~Bedeschi, \emph{{A detector concept proposal for a circular $e^+e^-$
  collider}}, \href{https://doi.org/10.22323/1.390.0819}{\emph{PoS} {\bfseries
  ICHEP2020} (2021) 819}.

\bibitem{Viazlo_2019}
O.~Viazlo, \emph{Fine-grained calorimeters for experiments at clic and fcc-ee},
  \href{https://doi.org/10.1088/1742-6596/1162/1/012015}{\emph{Journal of
  Physics: Conference Series} {\bfseries 1162} (2019) 012015}.

\bibitem{Bacchetta:2019fmz}
N.~Bacchetta et~al., \emph{{CLD -- A Detector Concept for the FCC-ee}},
  \href{https://arxiv.org/abs/1911.12230}{{\ttfamily 1911.12230}}.

\bibitem{ALLISON2016186}
{\scshape GEANT4} collaboration, \emph{Recent developments in geant4},
  \href{https://doi.org/10.1016/j.nima.2016.06.125}{\emph{Nuclear Instruments
  and Methods in Physics Research Section A: Accelerators, Spectrometers,
  Detectors and Associated Equipment} {\bfseries 835} (2016) 186–225}.

\bibitem{dd4hep}
M.~Frank, F.~Gaede, C.~Grefe and P.~Mato, \emph{{DD4hep: A Detector Description
  Toolkit for High Energy Physics Experiments}}, {\emph{J. Phys.: Conf. Ser.}
  {\bfseries 513} (2014) 022010}.

\bibitem{EJefficiency}
{ELJen Technology}, \emph{{General purpose EJ-200, EJ-204, EJ-208, EJ-212}},
  2023.

\bibitem{RMS90}
R.~Cassell, ``Rms vs rms90 comment.'' Note
  \url{https://agenda.linearcollider.org/event/2703/}, 2008.

\bibitem{Wenzel_2012}
H.~Wenzel, \emph{Simulation studies of a total absorption dual readout
  calorimeter},
  \href{https://doi.org/10.1088/1742-6596/404/1/012049}{\emph{Journal of
  Physics: Conference Series} {\bfseries 404} (2012) 012049}.

\bibitem{Akchurin:2014xpa}
N.~Akchurin et~al., \emph{{Lessons from Monte Carlo simulations of the
  performance of a dual-readout fiber calorimeter}},
  \href{https://doi.org/10.1016/j.nima.2014.05.121}{\emph{Nucl. Instrum. Meth.
  A} {\bfseries 762} (2014) 100}.

\bibitem{Chekanov:2014fga}
S.V.~Chekanov, \emph{{HepSim: a repository with predictions for high-energy
  physics experiments}}, \href{https://doi.org/10.1155/2015/136093}{\emph{Adv.
  High Energy Phys.} {\bfseries 2015} (2015) 136093}
  [\href{https://arxiv.org/abs/1403.1886}{{\ttfamily 1403.1886}}].

\bibitem{githubc}
S.V.~Chekanov and S.~Eno, ``{DualCrystalHcal repository}.''
  \url{https://github.com/chekanov/DualCrystalHcal}, 2023.

\end{thebibliography}\endgroup

\end{document}